\documentclass[10pt,journal,twocolumn]{IEEEtran} 
\usepackage{wrapfig,booktabs,fancyhdr,amsmath,amsfonts,tabularx}
\usepackage{cite,bm,bbm,amssymb,amsthm,url,multirow,times,enumitem,comment}
\usepackage{mathtools,siunitx,balance,tikz,adjustbox,array,graphicx,xspace}
\usepackage{mathrsfs}
\usepackage[font=footnotesize]{subcaption}
\usepackage[linesnumbered,ruled]{algorithm2e}
\usepackage[colorlinks=true, linkcolor=black, citecolor=blue, urlcolor=blue]{hyperref}
\newcommand{\vb}{\boldsymbol}

\newcommand{\sfrac}[2]{\textstyle \frac{#1}{#2}}
\newcommand{\ba}{\begin{array}}
\newcommand{\ea}{\end{array}}
\newcommand{\sinc}{\text{sinc}}
\newcommand{\define}{\triangleq}

\DeclareMathAlphabet{\mathpzc}{OT1}{pzc}{m}{it}

\newcommand{\Fc}[1]{\ensuremath{F_{\text{c}#1}}}
\newcommand{\Fs}{\ensuremath{F_\text{s}}}

\newcommand{\N}{\ensuremath{N}}

\newcommand{\Tsym}{\ensuremath{T_\text{sym}}}
\newcommand{\Ts}{\ensuremath{T_\text{s}}}

\newcommand{\Tf}{\ensuremath{T_\text{f}}}
\newcommand{\tf}{\ensuremath{t_\text{f}}}
\newcommand{\dtfp}{\ensuremath{{\delta t}^{\prime}_\text{f}}}
\newcommand{\dtrp}{\ensuremath{{\delta t}^{\prime}_\text{r}}}
\newcommand{\dtf}{\ensuremath{{\delta t}_\text{f}}}
\newcommand{\tc}{\ensuremath{t_\text{c}}}
\newcommand{\td}{\ensuremath{t_\text{d}}}
\newcommand{\sd}{\ensuremath{s_\text{d}}}
\newcommand{\tr}{\ensuremath{t_\text{r}}}
\newcommand{\wwr}{\ensuremath{w_\text{r}}}
\newcommand{\tfp}{\ensuremath{{t}^\prime_\text{f}}}
\newcommand{\trp}{\ensuremath{{t}^\prime_\text{r}}}
\newcommand{\ttr}{\ensuremath{\tilde{t}_\text{r}}}
\newcommand{\ttrp}{\ensuremath{{\tilde{t}}^{\prime}_\text{r}}}
\newcommand{\ttrpp}{\ensuremath{{\tilde{t}}^{\prime\prime}_\text{r}}}
\newcommand{\ttrppp}{\ensuremath{{\tilde{t}}^{\prime\prime\prime}_\text{r}}}
\newcommand{\Ml}{\ensuremath{\mathcal{M}_{l}}}
\newcommand{\Df}{\ensuremath{\Delta t_\text{f}}}
\newcommand{\Dc}{\ensuremath{\Delta t_\text{c}}}

\newcommand{\dTOF}{\ensuremath{\delta t_\text{tof}}}
\newcommand{\dTOFg}{\ensuremath{\delta t_\text{tofg}}}
\newcommand{\datm}{\ensuremath{\delta t_\text{atm}}}

\newcommand{\F}{\ensuremath{F}}
\newcommand{\Tg}{\ensuremath{T_\text{g}}}
\newcommand{\Ng}{\ensuremath{N_\text{g}}}
\newcommand{\Na}{\ensuremath{N_\text{a}}}
\newcommand{\Nag}{\ensuremath{N_\text{ag}}}
\newcommand{\Nbi}{\ensuremath{N_{\text{b}i}}}
\newcommand{\Nsi}{\ensuremath{N_{\text{s}i}}}
\newcommand{\Nasi}{\ensuremath{N_{\text{as}i}}}
\newcommand{\Tag}{\ensuremath{T_\text{ag}}}

\newcommand{\Ff}{\ensuremath{F_\text{f}}}
\newcommand{\Tfg}{\ensuremath{T_\text{fg}}}

\newcommand{\betaf}{\ensuremath{\beta_\text{f}}}

\begin{document}

\title{Timing Properties of the \\ Starlink Ku-Band Downlink}

\author{
  \IEEEauthorblockN{Wenkai Qin\IEEEauthorrefmark{1},
    Andrew M. Graff\IEEEauthorrefmark{2},
    Zachary L. Clements\IEEEauthorrefmark{1},
    Zacharias M. Komodromos\IEEEauthorrefmark{2},
    Todd E. Humphreys\IEEEauthorrefmark{1}} \\
  \IEEEauthorblockA{\IEEEauthorrefmark{1}\textit{Department of Aerospace
      Engineering and Engineering Mechanics, The University of Texas at Austin}} \\
  \IEEEauthorblockA{\IEEEauthorrefmark{2}\textit{Department of Electrical and
      Computer Engineering, The
    University of Texas at Austin}}
}
\maketitle

\begin{abstract}
  We develop signal capture and analysis techniques for precisely extracting and
  characterizing the frame timing of the Starlink constellation's Ku-band
  downlink transmissions.  The aim of this work is to determine whether Starlink
  frame timing has sufficient short-term stability to support pseudorange-based
  opportunistic positioning, navigation, and timing (PNT).  A second goal is to
  determine whether frame timing is disciplined to a common time scale such as
  GPS time.  Our analysis reveals several timing characteristics not previously
  known that carry strong implications for PNT.  On the favorable side, periods
  of ns-level jitter in frame arrival times across all satellite versions
  indicate that Starlink hardware is fundamentally capable of the short-term
  stability required to support GPS-like PNT.  But there are several unfavorable
  characteristics that, if not addressed, will make GPS-like PNT impractical:
  (1) The v1.0 and v1.5 Starlink satellites exhibit once-per-second abrupt frame
  timing adjustments whose magnitude (as large as 100s of ns) and sign appear
  unpredictable.  Similar discontinuities are also present in the v2.0-Mini
  frame timing, though smaller and irregularly spaced. (2) Episodic 15-s periods
  of high frame jitter routinely punctuate the nominal low-jitter frame arrival
  timing.  (3) Starlink frame timing is disciplined to GPS time, but only
  loosely: to within a few ms by adjustments occurring every 15 s; otherwise
  exhibiting drift that can exceed 20 ppm.  These unfavorable characteristics
  are essentially incompatible with accurate PNT.  Fortunately, they appear to
  be a consequence of software design choices, not hardware limitations.
  Moreover, they could be compensated with third-party-provided corrections.
\end{abstract}


\newif\ifpreprint
\preprintfalse

\ifpreprint

\pagestyle{plain}
\thispagestyle{fancy}  
\fancyhf{} 
\renewcommand{\headrulewidth}{0pt}
\rfoot{\footnotesize \bf January 2025 preprint of paper to be submitted for review} \lfoot{\footnotesize \bf
  Copyright \copyright~2025 by Wenkai Qin, Andrew M. Graff, Zachary L. Clements, \\
  Zacharias M. Komodromos, Andrew M. Graff, and Todd E. Humphreys}

\else

\thispagestyle{empty}
\pagestyle{plain}

\fi


\section{Introduction}
Global navigation satellite system technology (GNSS) is currently the most
prevalent used for positioning, navigation, and timing (PNT). However,
traditional GNSS is vulnerable to jamming and spoofing attacks that can leave
users without the ability to navigate or synchronize time
\cite{humphreysGNSShandbook}, and threats to traditional GNSS are multiplying
dramatically
\cite{murrian2021leo,clements2023PlansDirectGeo,opsgroup2024spoofing}.
According to OPSGROUP, an international association of air transport
professionals, GNSS spoofing incidents increased by 500\% from 2023 to 2024
\cite{opsgroup2024spoofing}.  To strengthen radionavigation, researchers have
recently focused on augmenting traditional GNSS with large low Earth orbit (LEO)
communications constellations \cite{stock2024survey}, with some proposing a
combined communications-PNT service for future constellations
\cite{iannucci2022fusedLeo}. Because these constellations offer higher power and
wider bandwidth, they are inherently resilient to adversarial
interference. Further, the two-way high-rate connectivity afforded by broadband
communications constellations enables desirable features such as user
authentication and near-zero age of ephemeris and clock models.

Researchers interested in a free-to-use radionavigation receiver have
investigated opportunistic approaches to PNT, i.e., PNT extraction with no
direct cooperation from the constellation operator and limited \emph{a priori}
knowledge regarding satellites' ephemerides and signals. SpaceX's Starlink
constellation is of particular interest: it offers the widest signal
availability, serving millions of subscribers worldwide with its 7,000+
satellites \cite{gomez2024starlink}. Opportunistic approaches using Starlink's
Ku-band signals (10.7-12.7 GHz) have already proven fruitful: researchers in
several groups have independently demonstrated Doppler-based positioning with
accuracy on the order of 10 m \cite{kassas2024adAstra, neinavaie2023cognitive,
  jardak2023practical, yang2023starlink, kozhaya2024fundamental,
  psiaki2021navigation, moore2024time}. Unfortunately, Doppler-based techniques
cannot approach the exquisite timing precision offered by traditional GNSS: even
in the optimistic scenarios posed in \cite{psiaki2021navigation, moore2024time},
timing accuracy is limited to no better than 0.1 ms. By contrast,
pseudorange-based PNT from Starlink holds the potential for both meter-accurate
positioning and ns-accurate timing \cite{iannucci2022fusedLeo,
  iannucci2020fused}. But whether Starlink signals could support precise
pseudorange-based PNT remains an open question whose answer depends on the
details of the broadcast signals, including modulation, timing, and spectral
characteristics.

The authors of \cite{humphreys2023starlinkSignalStructure} uncovered key
information regarding the Starlink downlink frame structure, synchronization
sequences, and spectral characteristics.  Follow-on work in
\cite{Komodromos2024WeakStarlinkSignal} and \cite{kozhaya2024trickOrTreat}
discovered other predictable elements of each frame, and \cite{qin2023turret}
revealed that Starlink beam switching occurs at 15-s, approximately-GPS-aligned
intervals.  Other studies have shown simulated impacts of various clock types in
LEO \cite{wang2022LEOclock}, or have developed methods for predicting LEO clock
corrections, such as those developed for the GRACE mission
\cite{ge2023characteristics, jiang4818891characteristics}. Nonetheless, no prior
work has characterized the stability of the Starlink frame timing, nor
investigated its precise relationship to an absolute time scale such as GPS time
(GPST), despite these essential details being prerequisite to development of
Starlink-based PNT, whether opportunistic or not.

Our paper closes this knowledge gap. We leverage the signal structure details
revealed in \cite{humphreys2023starlinkSignalStructure} to conduct a focused
study of Starlink timing properties. First, we evaluate the short-term frame
clock stability, defined as the set of clock behaviors that manifest on the
order of several seconds or less whose study indicates both the quality of
crystal oscillator(s) onboard the Starlink satellite vehicles (SVs) and the
predictability of any onboard clock corrections. Second, we investigate the
absolute frame timing characteristics; i.e., we determine the extent to which
Starlink frame clocks are disciplined to GPST.

A preliminary conference version of this paper appeared as
\cite{qin2024shortTermFrameTiming}. The current version extends the former with
a deeper examination of short-term frame clock stability, a new section
analyzing absolute frame timing, a presentation of theoretical measurement
bounds, and stronger conclusions.

\section{Signal Capture}
\label{sec:signal-capture}
While many particulars of our signal capture system match those presented in
\cite{humphreys2023starlinkSignalStructure}, we reiterate the major signal
pathways and note some updates made for the purposes of timing analysis.

\begin{figure}[t]
\centering
\includegraphics[width=\columnwidth]
{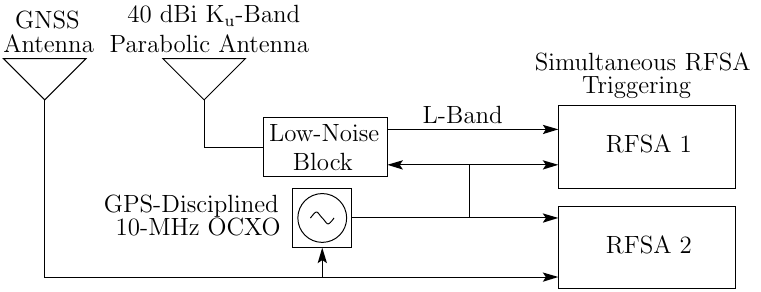}
\caption{Block diagram of the simultaneous Starlink and GPS signal capture system.}
\label{fig:captureSetup}
\end{figure}

Fig. \ref{fig:captureSetup} shows a block diagram of the signal capture
system. The Starlink signal chain begins with reception of Ku-band signals
through a 40 dBi parabolic dish antenna followed by frequency downconversion to
L-band and signal amplification in a low-noise block (LNB). Further
downconversion, amplification, filtering, and digital sampling at rates ranging
from 50.0 to 62.5 Msps occur within the radio-frequency signal analyzer (RFSA)
1. This path is equivalent to the narrowband capture chain in
\cite{humphreys2023starlinkSignalStructure}.

A parallel GPS L1 capture pathway comprises an active GNSS antenna connected to
an RFSA identical to RFSA 1, within which are realized the same signal
processing and sampling operations as for the Starlink signal pathway. This
second RFSA is denoted RFSA 2 in Fig.  \ref{fig:captureSetup}.  

A key feature of the parallel capture setup is the GPS-disciplined 10-MHz
oven-controlled crystal oscillator (OCXO), which drives both Starlink and GPS L1
signal pathways, allowing for synchronous capture with a unified clock model.
To verify this, we captured signals from the GNSS antenna simultaneously with
both RFSAs, splitting the signal through equal-length cables just before
insertion into each. Both data captures were then independently processed via
GRID, our laboratory's science-grade software-defined GNSS receiver
\cite{nichols2022launch, clements2021bitpackingIonGnss,pany2024historySdr}. By
this we found that the two RFSAs' sample trains are synchronous to better than 1
ns when configured with matching sampling rates, and offset by measurable
deterministic amounts when the sampling rates were not matched but were integer
multiples of one another.

To measure the system's cable delay from the Ku-band LNB to RFSA 1, required for
absolute timing analysis, we performed the same experiment but routed one branch
of the split signal in a loopback configuration to near the Ku-band antenna LNB
and back before insertion into RFSA 1. Dividing the observed differential
latency by two yielded the unknown cable delay. Finally, to measure the
additional cable delay in the GNSS capture pathway beyond that of the Starlink
pathway, we substituted a GNSS antenna for the Ku-band antenna and LNB and
measured the differential signal timing via simultaneous dual-RFSA capture and
post-processing through GRID.

With this setup, we can capture up to 62.5 MHz of real-time Starlink bandwidth
through RFSA 1 with highly stable uniform sampling, and mark each sample with a
GPST timestamp to ns accuracy.  This enables us to precisely determine the
arrival time of a Starlink frame in GPST and study the stability of arrival
times from frame to frame.

\section{Frame Structure and System Concepts}
\label{sec:sys-concept-term}
This section introduces the Starlink frame structure, drawn from
\cite{humphreys2023starlinkSignalStructure}, and additional concepts and
terminology needed to understand Starlink operation in general and
Starlink-based PNT in particular.

\subsection{Frame Structure}
\label{sec:frame-structure-1}

\begin{figure*}[t]
  \centering
  \includegraphics[width=0.75\textwidth]{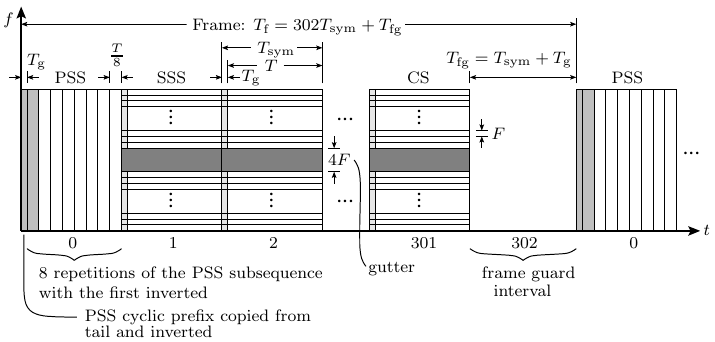}
  \caption{Frame layout for the Ku-band Starlink downlink along time-frequency
    dimensions, from \cite{humphreys2023starlinkSignalStructure}.  Indices along
    the horizontal axis enumerate the 303 intervals that constitute a single
    frame.  The quantity $T$ is the useful (non-cyclic) OFDM symbol interval,
    $\Tg$ is the symbol guard interval (cyclic prefix), and $F$ is the
    subcarrier spacing.  Other quantities are defined in the text.}
  \label{fig:frame-diagram}
\end{figure*}

As shown in Fig. \ref{fig:frame-diagram}, each Starlink downlink orthogonal
frequency division multiplexing (OFDM) frame consists of 302 intervals of length
$\Tsym = \SI{4.4}{\micro\second}$ plus a frame guard interval $\Tfg$, for a
total frame period of $\Tf = 1/\Ff$ s, where $\Ff = 750$ Hz is the frame rate.
Each frame begins with the primary synchronization sequence (PSS), which is
natively represented in the time domain, followed by the secondary
synchronization sequence (SSS), which is formatted as a standard 4QAM OFDM
symbol. The final occupied symbol interval in each frame is the coda symbol
(CS), which is followed by the frame guard interval. Subsequent frames may be
present or not, depending on user demand.

The time-domain modulation sequence of the PSS, and the frequency-domain symbol
sequence of the SSS were revealed in
\cite{humphreys2023starlinkSignalStructure}. These sequences allow a Starlink
receiver to perform the channel estimation necessary to demodulate each OFDM
symbol in the frame. In particular, the PSS allows the receiver to precisely
identify the beginning of each frame, while the SSS allows it to perform
equalization across all subcarriers. The PSS and SSS have a fixed phase
relationship, which allows coherent integration across the combined PSS + SSS
interval, permitting more accurate synchronization and carrier frequency offset
(CFO) estimation, as will be further examined in Section \ref{sec:error_var}.

The role of the CS, called the CSS in
\cite{humphreys2023starlinkSignalStructure}, is not entirely clear. Its
frequency-domain symbol sequence is highly predictable, but not entirely so,
unlike the SSS \cite{Komodromos2024WeakStarlinkSignal}. Likewise, other OFDM
symbols in each frame have varying degrees of predictability, whether due to
pilot symbols meant to aid channel estimation or to default symbols transmitted
when user data do not occupy all the payload-bearing symbols in a frame
\cite{Komodromos2024WeakStarlinkSignal, kozhaya2024trickOrTreat}. For the timing
analysis in this paper, the local signal replica used for correlation comprises
only the coherent PSS + SSS combination so that, insofar as possible, results
are uniform across all received frames.

\subsection{Assigned Beams}
At any given time, a user within Starlink's $\pm 53^\circ$ latitude primary
coverage area may have a direct line of sight to dozens of overhead SVs above
$20^\circ$ elevation \cite{iannucci2022fusedLeo}.  Each Starlink SV is capable
of simultaneously directing up to 48 downlink beams to terrestrial service
cells, 16 for each of its three downlink phased arrays. A proprietary beam
assignment procedure assigns each beam to a $\sim$20-km-diameter service cell
\cite{pekhterev2021bandwidth}.  A service cell may be illuminated by up to 16
beams simultaneously, two beams for each of the eight frequency channels
identified in \cite{humphreys2023starlinkSignalStructure}.  We call the $\Nbi$
beams deliberately directed to the $i$th service cell \emph{assigned beams}.

The number of distinct SVs casting assigned beams onto the $i$th service cell,
$\Nasi \leq \Nbi \leq 16$, may be less than $\Nbi$ because a given SV may
project multiple beams onto the same cell, each on a different channel or with a
different polarization (right-hand vs. left-hand circular polarization).  As
Starlink user density has increased over the past few years, $\Nasi$ and $\Nbi$
have generally increased.  It was shown in \cite{qin2023turret} that a search
procedure prioritizing high-elevation SVs was effective at finding multiple
illuminating SVs.

\subsection{Fixed Assignment Interval}
Several publications have noted the existence of a 15-s Starlink network
reconfiguration interval \cite{qin2023turret, huston2024transportProtocol,
  mohan2024multifaceted, grayver2024position}. Because all beam assignments
remain fixed over the duration of this interval, we call it the \emph{fixed
  assignment interval} (FAI). FAI boundaries appear to be approximately aligned
with GPST \cite{qin2023turret}. As discussed later, noticeable transitions in
Starlink frame timing behavior occur at FAI boundaries. Accordingly, we split
captures along FAI boundaries when analyzing frame clock behaviors.

We write $\Nbi(l)$ to indicate the number of assigned beams for the $i$th cell
during the $l$th FAI, and $\Nasi(l) \leq \Nbi(l) \leq 16$ to indicate the number
of distinct SVs casting assigned beams onto the $i$th cell during the $l$th FAI.

\subsection{Side Beams}
\begin{figure}[t]
\centering
\includegraphics[width=\columnwidth]
{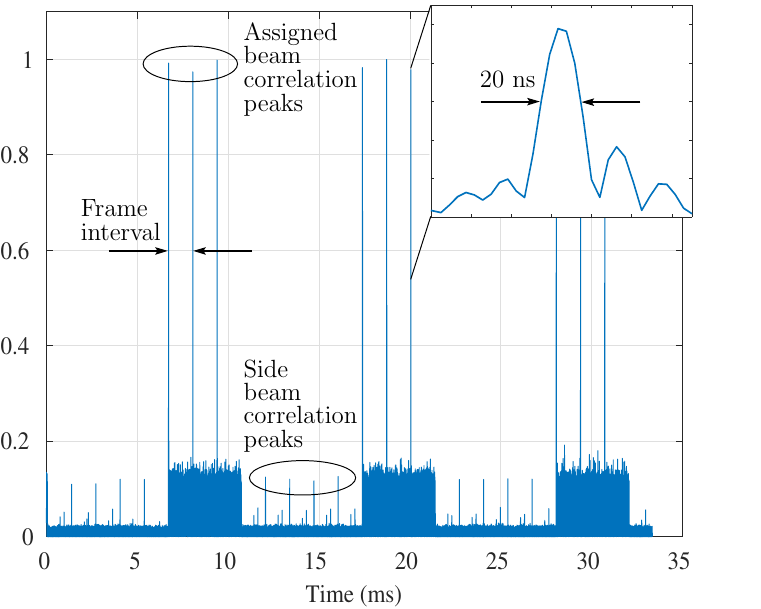}
\caption{Normalized cross-correlation of received Starlink data against a local
  PSS + SSS replica yields sharp peaks at the beginning of each frame.  The
  peaks' primary lobe is approximately 20 ns wide for a 55-MHz-bandwidth
  capture.  The nominal interval between frames is $\Tf = 1/750$ s. The data
  shown are for STARLINK-30580, a v2.0-Mini SV, from signals captured in
  February 2024, but the frame correlation pattern is broadly representative of
  signals captured since 2022 for v1.0, v1.5, and v2.0-Mini SVs.}
\label{fig:frameClockExamplePlotP}
\end{figure}

Assigned beams are not the only ones from which a receiver in a given service
cell may extract PNT information.  Consider
Fig. \ref{fig:frameClockExamplePlotP}, which shows the normalized
cross-correlation of a short interval of received Starlink data against a local
signal replica composed of a coherent PSS + SSS combination.  The highest
correlation peaks shown correspond to frames from an assigned beam, whereas the
weaker peaks correspond to frames from \emph{side beams}, or beams directed
toward other service cells.  The assigned beam's pre-correlation SNR over this
interval is approximately 21 dB, whereas for the side beam corresponding to the
lower circled correlation peaks the pre-correlation SNR is only 3 dB.  Despite
their lower SNR, it is possible to extract time of arrival (TOA) measurements
from such side-beam peaks precisely enough to support GPS-like PNT.

The post-correlation SNR of side-beam correlation peaks depends on the side
beam's pre-correlation SNR and on the processing gain afforded by correlation
against known synchronization sequences or other predictable elements of a
frame. Reference \cite{komodromos2023IONsimulator} shows that correlation
against a coherent PSS + SSS replica can support reliable signal detection
across a large blind Doppler and time offset search space at a low false alarm
rate provided that frames have a pre-correlation SNR higher than -18 dB.  If the
receiver can somehow assemble a local replica of an entire frame, whether
because a Starlink SV transmits a known default frame during a time of low user
demand \cite{Komodromos2024WeakStarlinkSignal}, or because a third party
receives the frame through a high-gain antenna and provides its contents through
a side channel to the receiver, the full-frame processing gain makes it possible
to reduce the detection threshold for pre-correlation SNR to -40 dB
\cite{komodromos2023IONsimulator}.  A later section will examine frame TOA
measurement precision as a function of pre-correlation SNR and processing gain.
For now, suffice it to say that useful TOA measurements could be extracted from
all the minor peaks visible in Fig. \ref{fig:frameClockExamplePlotP}.

There are two possible scenarios under which side beams may be observed. First,
the side lobes of a beam may be strong enough for the beam to be detectable by a
receiver in a service cell different from the one to which the beam is assigned.
Signals received in this way may be severely attenuated: for the Ku-band
communications SVs studied in \cite{hills2023controlling}, the downlink phased
array's side lobes were attenuated by $\sim$30 dB relative to the main lobe.

Second, spillover from a beam assigned to a cell near the one occupied by the
receiver can yield a detectable side beam. A beam's main lobe creates a ground
coverage region whose shape can be modeled as the elliptical intersection of a
cone aligned with the main lobe's boresight and Earth's surface
\cite{hills2022feasibility}. Let $R_{\text{b}i}$ represent, for a given
processing gain and for a beam assigned to the $i$th service cell, the region
within which a receiver may reliably extract precise TOA measurements, and let
$R_{\text{c}i}$ be the $i$th hexagonal Starlink service cell.  For reliable
broadband service, it must be the case that
$R_{\text{c}i} \subset R_{\text{b}i}$. Moreover, recognizing that the SNR
requirement for TOA measurement is much lower than that for broadband
communications, one expects the spillover region
$R_{\text{so}i} \define R_{\text{b}i} \setminus R_{\text{c}i}$ to be large. This
is especially so given that $R_{\text{b}i}$'s elliptical eccentricity varies
inversely with the assigned beam's elevation angle viewed from its target
service cell \cite{hills2022feasibility}, and given that Starlink seeks to lower
its minimum transmission elevation angle from $25^\circ$ to $20^\circ$
\cite{brodkinStarlinkGigabit}.  Thus, spillover-induced side beams are expected
to be plentiful.

We denote by $\Nsi(l) \geq \Nasi(l)$ the total number of SVs from which a
receiver in the $i$th service cell may reliably extract precise TOA measurements
during the $l$th FAI.  Due to the presence of side beams, $\Nsi(l)$ may be
significantly larger than $\Nasi(l)$.  For Starlink-based PNT, larger $\Nsi(l)$
ensures solution robustness and tends to increase the geometric diversity of a
navigation solution, which decreases the solution's dilution of precision
\cite{odijkGNSShandbook}.

\subsection{Signal Terminology}
Captured signals may include simultaneous transmissions from both assigned and
side beams. We define a \emph{composite signal} as one with significant power
contributions from multiple beams, whereas a \emph{simplex signal} contains
significant power only from a single beam.  Within a composite signal capture,
the strongest transmission from a single beam is referred to as the
\emph{dominant signal}, while weaker transmissions from other beams are called
\emph{secondary signals}. Typically, the dominant signal corresponds to an
assigned beam.

We identify a particular sequence of correlation peaks during a given FAI as
corresponding to a unique beam by a two-factor test: (1) the peaks have
approximately the same magnitude, and (2) the peaks are spaced from one another
by integer multiples of $\Tf = 1/\Ff = 1/750$ s according to the relevant frame
clock. Given a sufficiently long capture within a single FAI, this test is
adequate to associate all correlation sequences of significant magnitude with
their respective beams.  Thus, for example, the largest 9 peaks in Fig.
\ref{fig:frameClockExamplePlotP} correspond to frames from a unique assigned
beam, and the set of 12 lower peaks, including the four lower circled ones,
correspond to frames from a unique side beam.

To classify single-beam signals as dominant or secondary, we estimate their SNR
by comparing post-correlation peaks with a noise floor estimate, derived from
the complex signal variance in intervals presumed to lack active transmissions.
However, strictly speaking, it is not always possible to ensure the noise floor
estimate is completely free of secondary signals.

Whether a given captured signal is composite or simplex depends on the
processing gain afforded by the local signal replica. For a coherent PSS + SSS
replica, as in Fig. \ref{fig:frameClockExamplePlotP}, signals with
pre-correlation SNR below -18 dB are effectively undetectable
\cite{komodromos2023IONsimulator}, rendering them insignificant as secondary
signals.

\section{Clock Models}
\label{sec:clock-models}
As with any analysis of PNT systems based on radio wave propagation,
unambiguous models of the various clocks involved are key to understanding and
characterizing the system. 

\begin{figure}[t]
\centering
\includegraphics[width=\columnwidth]{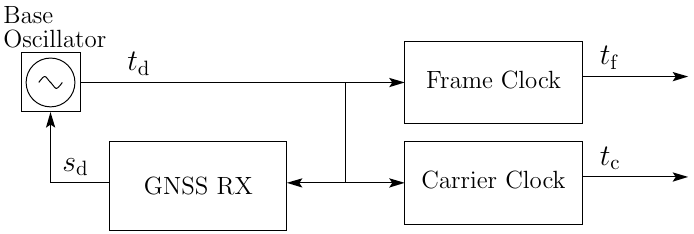}
\caption{Clock cascade model for a single Starlink downlink beam.}
\label{fig:clock-block-diag}
\end{figure}

\subsection{Beam-Specific Clock Cascade Model}
\label{sec:beam-specific-clock}
Fig. \ref{fig:clock-block-diag} presents a clock cascade model for a single
Starlink downlink beam.  The model is beam-specific because, as will be shown
later on, frame timing behavior may differ from beam to beam over the same FAI
for the same SV.  This remarkable observation implies that Starlink signal
timing is much different from that of traditional GNSS, in which a single clock
governs the whole transmission cascade across all frequencies.  To be sure, in
traditional GNSS timing offsets may be present between different spreading codes
on the same carrier and across multiple carriers, resulting in so-called
differential code biases \cite{montenbruck2014differential}.  But the code rates
as transmitted are all a constant multiple of a single base clock's reference
frequency.  By contrast, the frame sequences on different beams from the same
Starlink SV may differ in both time offset and rate, a fact with significant
implications for Starlink-based PNT.

The root of each clock cascade is a base oscillator.  As will be shown, this
oscillator is shared across all beams from a given SV.  An onboard GNSS receiver
driven by the base oscillator produces a clock disciplining signal $\sd$ that
corrects the base oscillator towards GPST. Together, the base oscillator and the
GNSS receiver form a closed feedback loop that produces the GNSS-disciplined
timing signal $\td$.

\subsection{Frame and Carrier Clocks}
\label{sec:frame-carrier-clocks}
The GNSS-disciplined timing signal $\td$ drives the frame and carrier clocks. In
turn, the frame clock's signal $\tf$ governs the timing of frames transmitted by
the SV, and the carrier clock's signal $\tc$ drives the carrier onto which the
information-bearing frames are modulated.  The frame clock is likely a
software-based clock whose output $\tf$ depends on when baseband frames are
loaded into buffers for mixing to radio frequency (RF) and subsequent
transmission.  The carrier clock is likely transparent, meaning that
$\tc = \td$, but it is represented in a manner identical to the frame clock for
full model generality.

Note that in traditional GNSS the clock driving each SV's spreading code
(analogous to the frame clock) also drives the underlying carrier that the
spreading code modulates \cite{meurerAndAntreichGNSShandbookSignals}.  Thus, the
code and the carrier---as transmitted---are locked together such that the code
chipping rate is a constant rational multiple of the carrier frequency.  One of
this paper's key findings, elaborated later, is that such is not the case for
the Starlink Ku-band downlink signals.  Instead, the frame and carrier clocks
operate somewhat independently, which is why they are represented separately in
Fig. \ref{fig:clock-block-diag}.

We represent frame and carrier clock offsets from $\td$ as $\Df$ and
$\Dc$. These are related to $\td$, $\tf$, and $\tc$ by 
\begin{align}
  \label{eq:frameOffsetFrom_td}
  \td(t) &= \tf(t) - \Df(t) \\
  \label{eq:carrierOffsetFrom_td}
  \td(t) &= \tc(t) - \Dc(t)
\end{align}
where $t$ represents true time, or time according to an ideal clock, such as is
closely realized by GPST \cite{jekeliAndMontenbruckGNSShandbookTimeAndRefSys}.
In this paper, true time and GPST are taken to be synonymous.  The frame and
carrier clocks are related to $t$ by
\begin{align}
  t &= \tf(t) - \dtf(t) \\
  t &= \tc(t) - \delta\tc(t)
\end{align}
where $\dtf(t)$ and $\delta\tc(t)$ are the frame and carrier clock
offsets. 

The time derivatives of $\dtf(t)$ and of $\delta\tc(t)$ are called the
frame and carrier clock drift.  They are equivalent to the instantaneous
fractional frequency deviation, written generically as $y(t)$, on which clock
stability analysis is based \cite{beardAndSeniorGNSShandbook}.

\begin{figure}[t]
\centering
\includegraphics[width=\columnwidth]
{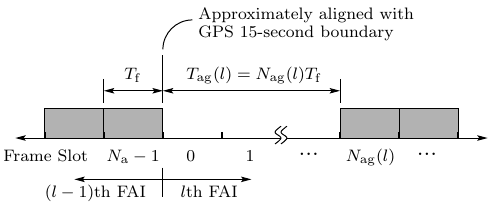}
\caption{Frame sequence timing diagram showing the transition from the $(l-1)$th FAI
to the $l$th FAI.}
\label{fig:frameTimeDiagram}
\end{figure}

\subsection{Discrete-Time Frame Clock}
The frame sequence timing diagram in Fig. \ref{fig:frameTimeDiagram} offers
further details about the frame clock.  Each frame as transmitted has a duration
of exactly $\Tf$ according to the frame clock.  Within each FAI, the frame slot
index increments from $0$ to $\Na - 1$, with $\Na = 11250$ being the number of
frame slots in one FAI.  Each FAI starts at the beginning of frame slot 0 and
lasts $\Na \Tf = 15$ s.  The interval of unoccupied frame slots at the beginning
of the $l$th FAI, called the FAI guard interval $\Tag(l) = \Nag(l) \Tf$, spans a
variable number of frame slots $\Nag(l)$.  Note that, for any FAI index $l$,
frame slot $\Nag(l)$ is occupied by definition, but other frame slots may not be
occupied.

Let $\tf(l,m)$ be the frame clock time at the instant when the frame in the
$m$th frame slot of the $l$th FAI begins to pass through the phase center of the
SV's downlink antenna, where $l$ and $m$ are zero-based indices. By definition,
we take this to be
\begin{align}
  \label{eq:tflm-def}
  \tf(l,m) \define 15l +  m\Tf 
\end{align}
Let $t^*(l,m)$ and $\dtf(l,m)$ be the corresponding GPST and frame
clock offset.  Then
\begin{align}
  \label{eq:tflm-def-true}
t^*(l,m) = \tf(l,m) - \dtf(l,m)
\end{align}
Another of this paper's key findings is that $\dtf(l,0) \approx 0$.  Stated
differently, a Starlink SV's frame clock departure from GPST at the
beginning of each FAI is small---typically less than a few ms.

\subsection{Receiver Clock}
Now consider the clock of a receiver tracking signals from the Starlink
downlink.  The receiver clock time $\tr$ is related to true time $t$ by
\begin{equation}
  \label{eq:receiverClock}
  t = \tr(t) - \delta \tr(\tr)
\end{equation}
The receiver clock offset $\delta\tr(\tr)$ is represented as a function of $\tr$
because it is natively ordered in receiver time in the course of solving for a
position and time solution.  The time derivative of $\delta \tr$ with respect to
$t$, denoted $\dot{\delta\tr}(\tr(t))$, is called the receiver clock drift.

Let $\tr(l,m)$ be the time of reception, according to the receiver clock, of the
frame transmitted at true time $t^*(l,m)$.  Let $\delta\tr(l,m)$ be the
corresponding receiver clock offset and $t_*(l,m)$ be the corresponding true
time of reception.  More precisely, $\tr(l,m)$ is the receiver clock time at
which the frame transmitted at true time $t^*(l,m)$ from the satellite's
downlink antenna's phase center first reached the receiver antenna's phase
center.  The receipt time $\tr(l,m)$ can be related to $t_*(l,m), t^*(l,m)$, and
$\tf(l,m)$ by
\begin{align}
  t_*(l,m) &= \tr(l,m) - \delta\tr(l,m)                               \label{eq:tgt-rx} \\
  t^*(l,m) &= \tr(l,m) - \delta\tr(l,m) - \dTOF(l,m)                  \label{eq:tgt-tx} \\
  \tf(l,m) &= \tr(l,m) - \delta\tr(l,m) - \dTOF(l,m) + \dtf(l,m) \label{eq:tflm-comp}
\end{align}
where $\dTOF(l,m)$ is the frame's time of flight from transmission to reception,
expressed as an interval in true time.

\section{Measurement Error Bounds}
\label{sec:error_var}
Here we provide theoretical bounds on the Doppler and TOA error variance based
on channel estimation using the PSS + SSS. We do this both to illustrate the
level of precision possible with Starlink-based PNT and to prepare for later
analysis investigating the relationship between frame and carrier Doppler.

\subsection{Doppler Error Variance}
\label{sec:dop_bounds}
A lower bound on Doppler frequency error variance for a signal $s(t)$ with
energy $E$, noise spectral density $N_0$, and mean time epoch $t_0$ can be
expressed in the form \cite[Eq. 11.27]{SkolnikRadarSystems}
\begin{align}
  \sigma_{f_\text{D}}^2 \geq \frac{1}{2\alpha^2 \sfrac{E}{N_0}}
\end{align}
where $\alpha^2$ is the effective squared time duration, defined as
\begin{align}
  \alpha^2 \define
  \frac{\int_{-\infty}^{\infty}(2\pi (t-t_0))^2 |s(t)|^2 dt}{\int_{-\infty}^{\infty} |s(t)|^2 dt}
\end{align}
The PSS signal given in \cite[Eq. 34]{humphreys2023starlinkSignalStructure} has
an effective time duration of
$\alpha_{\text{PSS}} = \SI{7.745}{\micro\second}$, while the SSS signal given
in \cite[Eq. 37]{humphreys2023starlinkSignalStructure} has an effective time
duration of $\alpha_{\text{SSS}} = \SI{7.8428}{\micro\second}$. The coherent
combination of PSS and SSS in a single frame results in
$\alpha_{1} = \SI{15.975}{\micro\second}$, and the coherent combination across
two frames results in $\alpha_{2} = \SI{4.2}{\milli\second}$.

\begin{figure}[t]
  \centering
  \includegraphics[width=0.9\linewidth]{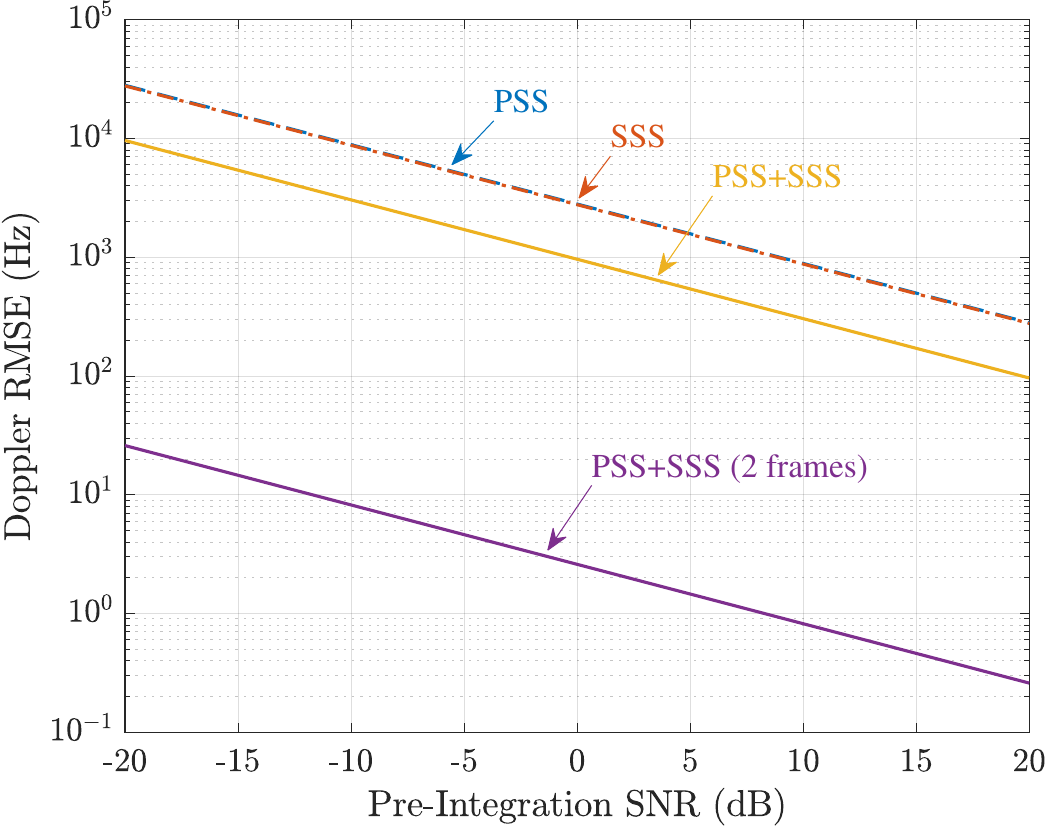}
  \caption{Doppler frequency RMSE bounds for various combinations of coherent processing based on the PSS and SSS.}
  \label{fig:doppler_bound}
\end{figure}

Fig. \ref{fig:doppler_bound} plots the Doppler frequency root-mean-square error
(RMSE) bound in units of Hz for the PSS-only, SSS-only, single-frame PSS + SSS,
and two-frame PSS + SSS signals (i.e., coherent correlation across the PSS + SSS
combination in a pair of adjacent frames). One may conclude from
Fig. \ref{fig:doppler_bound} that a Doppler estimation based on a single
coherent PSS + SSS interval yields an RMSE far worse than assumed in
\cite{psiaki2021navigation}. Clearly, for Doppler-based Starlink PNT, coherent
processing across many more than two OFDM symbols will be required.

\subsection{Time of Arrival Measurement Error Variance}
Similar to the foregoing Doppler frequency bound, a lower bound
on TOA error variance for a signal with Fourier transform $S(f)$, energy $E$,
and noise spectral density $N_0$ can be expressed in the form \cite[Eq.
11.15]{SkolnikRadarSystems}
\begin{align}
  \sigma_{\tau}^2 \geq \frac{1}{2\gamma^2 \sfrac{E}{N_0}}
\end{align}
where $\gamma^2$ is the effective squared bandwidth defined as
\begin{align}
  \gamma^2 \define \frac{\int_{-\infty}^{\infty}(2\pi f)^2 |S(f)|^2 df}{\int_{-\infty}^{\infty} |S(f)|^2 df}.
\end{align}

The PSS signal given in \cite[Eq. 34]{humphreys2023starlinkSignalStructure}
has a Fourier transform
\begin{align}
  S_{\text{PSS}}(f) \define \frac{1}{\Fs}\sum_{k=-\Ng}^{\N-1} \text{rect}\left(\sfrac{f}{\Fs}\right)\exp{\left(-j2\pi (k+\Ng) \sfrac{f}{\Fs}\right)} p_k
\end{align}
and an effective bandwidth of $\gamma_{\text{PSS}} = \SI{4.3482e8}{\hertz}$.
Similarly, the SSS signal given in \cite[Eq.
1.37]{humphreys2023starlinkSignalStructure} has a Fourier transform
\begin{align}
  S_{\text{SSS}}(f) & \define \sum_{k=-\sfrac{\N}{2}}^{\sfrac{\N}{2}-1} X_{m1k}\exp{\left(-j2\pi \F \Tg k\right)}  G_{\text{s}}(f-\F k)\\
  G_{\text{s}}(f) & \define \mathscr{F}[g_{\text{s}}(t)] = \Tsym \sinc\left(\Tsym f\right)\exp\left(-j\pi\Tsym f\right)
\end{align}
and an effective bandwidth of $\gamma_{\text{SSS}} = \SI{4.3658e8}{\hertz}$.
The PSS and SSS signal combined in a single frame have a Fourier
transform of $S_{1}(f) = S_{\text{PSS}}(f) + \exp\left(-j2\pi \Tsym f\right)
S_{\text{SSS}}(f)$. In the case of coherent processing of the PSS + SSS across two frames, this becomes $S_{2}(f) = S_{1}(f)
+ \exp\left(-j2\pi \Tf f\right)S_{1}(f)$.

\begin{figure}[t]
  \centering
  \includegraphics[width=0.9\linewidth]{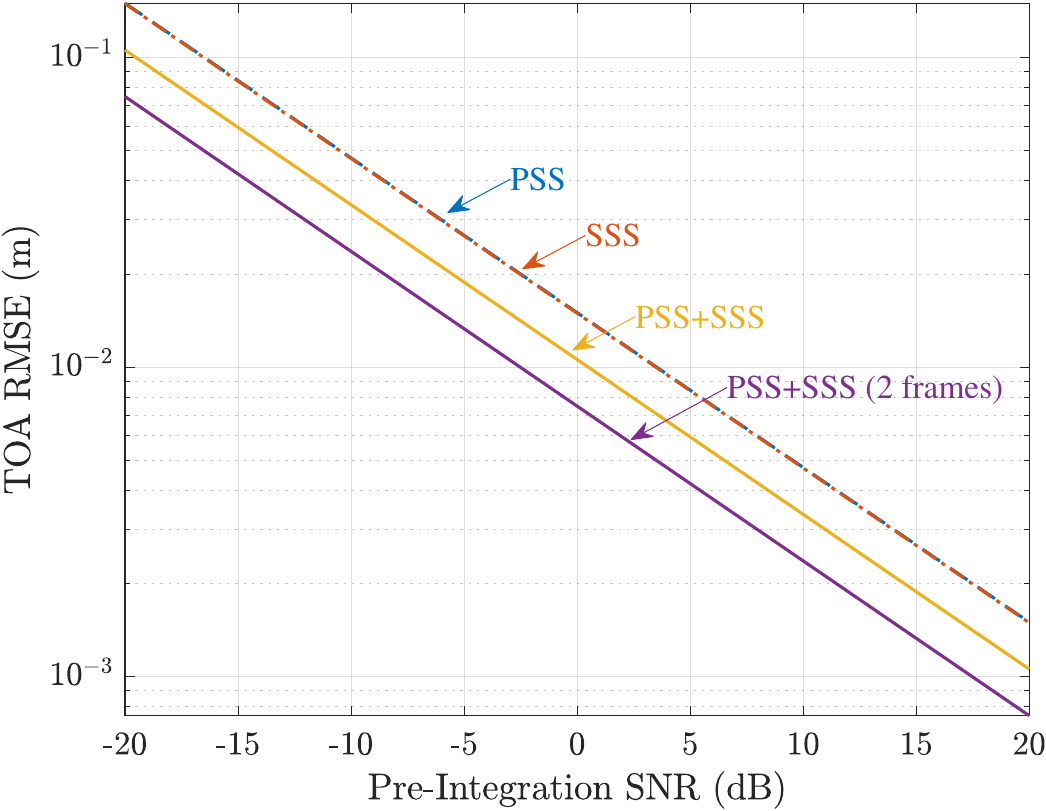}
  \caption{TOA RMSE bounds for various combinations of coherent processing based
    on the PSS and SSS.}
  \label{fig:toa_bound}
\end{figure}

Fig. \ref{fig:toa_bound} plots the TOA RMSE bound in units of meters for the
PSS-only, SSS-only, single-frame PSS + SSS, and two-frame PSS + SSS signals. Due
to the large bandwidth of the synchronization sequences, sub-centimeter accuracy
is theoretically achievable at moderate SNR.

\section{Data Preprocessing}
\label{sec:data-preproc}
We capture Starlink signals through RFSA 1 as shown in
Fig. \ref{fig:captureSetup} at a complex sampling rate ranging from 50.0 to 62.5
MHz.  Captured signals are then upsampled to the Starlink information symbol
rate $\Fs = 240$ MHz so that all subsequent operations may proceed as if the
full 240-MHz Starlink channel bandwidth had been captured.  The upsampled signal
is cross-correlated against a local signal replica consisting of the coherent
PSS + SSS combination to produce correlation peaks like those shown in
Fig. \ref{fig:frameClockExamplePlotP}.  For the $m$th frame, let
\begin{equation}
\label{eq:coherentCombPssSss}
x_{m01}(t) \define \begin{cases}
  x_{m0}(t), & 0 \leq t < \Tsym \\
  x_{m1}(t - \Tsym), & \Tsym \leq t < 2\Tsym \\
  0, & \mbox{otherwise}
\end{cases}
\end{equation}
be the coherent concatenation of the time-domain PSS and SSS functions from
\cite{humphreys2023starlinkSignalStructure}.  Because the PSS and SSS are
present in all Starlink downlink frames and are identical for all frames
$m \in \mathbb{Z}$ and all SVs, $x_{m01}(t)$ can be used for correlation against
all captured data.

The full discrete-time local signal replica used for correlation is the product
of $x_{m01}(t)$ and a complex exponential:
\begin{align}
  \label{eq:localReplicaWithDoppler}
  y_{m01}[k] \define x_{m01}(k\Ts(1-\beta))\exp\left(j2\pi\left[\Fc{}(1-\beta) -
  \bar{\Fc{}}\right]k\Ts\right)
\end{align}
Here, $\Ts = 1/\Fs$ is the sampling interval, $\Fc{}$ is the center frequency of
the OFDM channel, $\beta$ is the CFO parameter, and $\bar{\Fc{}} \approx \Fc{}$
is the center frequency to which the receiver is tuned.  Correlation proceeds in
blocks of between 30 and 60 frame intervals $\Tf$.  Over each block, a constant
$\beta$ is applied that maximizes correlation peak magnitudes within the block.
This process amounts to batch sequential frequency tracking.

A high-precision TOA measurement is extracted from the correlation peak for each
frame received. Unless otherwise noted, all frame timing measurements were
extracted from dominant signals with pre-correlation SNR exceeding -1 dB, which,
upon correlation against the replica signal in
(\ref{eq:localReplicaWithDoppler}), yields a post-correlation SNR of at least 25
dB for a $\geq 50$-MHz captured bandwidth \cite{komodromos2023IONsimulator}. Let
$\Ml$ be the set of occupied-frame indices for the $l$th FAI.  A sequence of
frame TOA measurements corresponding to $\Ml$ is extracted from the
cross-correlation function.  These are modeled as
\begin{equation}
  \label{eq:peakMeasurementModel}
\ttr(l,m) = \tr(l,m) + \wwr(l,m), \quad m \in \Ml
\end{equation}
where $\wwr(l,m)$ is zero-mean Gaussian measurement error with variance
$\sigma_w^2(l,m)$. The $m$th measurement $\ttr(l,m)$ is the receiver time of
the discrete sample instant at which the correlation peak for the $m$th frame is
maximized.  As can be appreciated by examining the zoomed inset in Fig. 3, the
measurement errors $\{\wwr(l,m) ~|~ m \in \Ml \}$ contain errors due to (1)
nearest-sample quantization of the maximizing location, (2) thermal noise, and
(3) peak rounding caused by filtering in the RFSA for the relatively narrow
bandwidth captured (e.g., 55 MHz for Fig.  \ref{fig:frameClockExamplePlotP} as
compared to the full $\Fs = 240$-MHz bandwidth). Their empirical standard
deviation is approximately $\sigma_w = 2$ ns, or 0.6 m in equivalent
length, which is well above the single-frame PSS + SSS RMSE bound for SNR
$\geq -1$ in Fig. \ref{fig:toa_bound}.  Nonetheless, this level of measurement
precision is more than adequate to study the short- and long-term Starlink frame
clock behavior because, as will be shown, variations in $\dtf(t)$ are typically
much larger than 2 ns.

\section{Short-Term Frame Clock Stability}
\label{sec:relat-frame-timing}
This section explores short term (within a single FAI) Starlink frame clock
behavior.  The emphasis here is on relative timing and high-frequency variations
in the frame clock offset $\dtf$, as opposed to absolute timing and
low-frequency variations.  Rearranging (\ref{eq:tflm-comp}) to isolate
$\dtf(l,m)$ yields
\begin{equation}
  \label{eq:isolated_df}
  \dtf(l,m) = \tf(l,m) - \tr(l,m) + \delta\tr(l,m) + \dTOF(l,m)
\end{equation}
Let $\dtfp(l,m)$ be equivalent to $\dtf(l,m)$ but with a $3$rd-order polynomial
fit across all $m \in \Ml$ removed. Assume like notation for the other terms in
(\ref{eq:isolated_df}).  Analysis of $\dtfp(l,m)$ is sufficient to characterize
the short-term properties of the frame clock because $\dtfp(l,m)$ retains the
high-frequency variations present in $\dtf(l,m)$. The detrended $\dtf(l,m)$ can
be modeled as
\begin{align}
  \label{eq:isolated_df_removed}
  \dtfp(l,m) &= \tfp(l,m) - \trp(l,m) + \delta\trp(l,m) +
               {\delta t}^\prime_\text{tof}(l,m) \nonumber \\
             &\approx -\trp(l,m) \\
             &\approx -\ttrp(l,m) \nonumber
\end{align}
This approximation is explained as follows: For $m \in \Ml$, $\tf(l,m)$ is an
affine function of $m$, which implies $\tfp(l,m) = 0$.  Likewise,
$\dtrp(l,m) \approx 0$ because the receiver clock is a GPS-disciplined OCXO with
negligible frequency error.  Finally,
${\delta t}^\prime_\text{tof}(l,m) \approx 0$ because the time of flight to an
SV in LEO can be modeled to better than 1 ns over a 15-s FAI as a 3rd-order
function.

In summary, for purposes of a short-term frame clock stability analysis,
$\dtfp(l,m)$ is a valid proxy for $\dtf(l,m)$, and $\ttrp(l,m)$, the
3rd-order-polynomial-detrended version of the frame TOA measurement
$\tilde{t}_\text{r}(l,m)$, is equivalent to $\dtfp(l,m)$ to within a sign
reversal and the ns-level measurement error $\wwr(l,m)$.

Note that until near the end of this section, we limit our analysis to v1.0 and
v1.5 SVs, as their frame clocks behave differently from those of v2.0-Mini SVs.

\begin{figure}[t]
\centering
\includegraphics[width=9.2cm]
{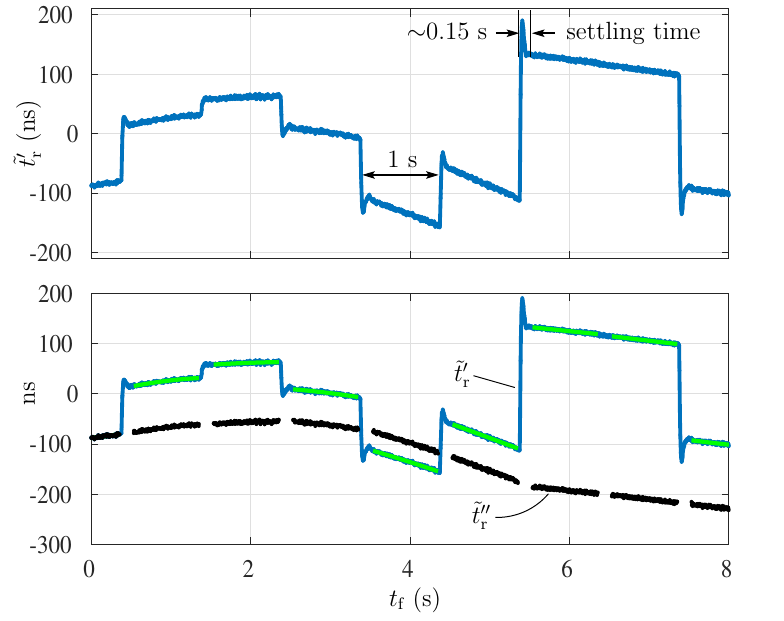}
\caption{Top: $\ttrp$ vs. $\tf$ for an example capture occurring within a single
  FAI. Bottom: $\ttrp$ (blue), piecewise 2nd-order polynomial fits to the
  truncated inter-adjustment segments (green), and $\ttrpp$, truncated
  inter-adjustment segments with first-order discontinuity eliminated
  (black). The data shown are for STARLINK-5141, a v1.5 SV, from signals
  captured in November 2023.}
\label{fig:frameTimeSmoothingP}
\end{figure}

\subsection{Periodic Frame Clock Adjustments}
\label{sub:nom-behavior}
The top panel of Fig. \ref{fig:frameTimeSmoothingP} shows an example $\ttrp$
trace for an 8-s interval within a single FAI.  Interpreting $\ttrp$ as a proxy
for the detrended frame clock deviation $\dtfp$, it is clear that the Starlink
frame clock exhibits abrupt adjustments at a regular 1-Hz cadence.  We believe
these are the result of coarse GNSS disciplining of the base oscillator.
Similar adjustments at the same 1-Hz cadence were evident in every capture of
signals from Starlink v1.0 and v1.5 SVs.  If due to GNSS disciplining, such
large and frequent adjustments---up to several hundred ns at 1 Hz---reflect a
base oscillator with poor stability.

Abrupt and coarse adjustments to the frame clock are obviously undesirable for
pseudorange-based PNT.  Unless they can be modeled or eliminated by some
differential scheme, such adjustments would cause large errors in pseudorange
modeling, and thus in position and timing estimation.  They act, in effect, like
the clock dithering implemented to intentionally degrade GPS accuracy under the
Selective Availability program discontinued in May 2000
\cite{odijkGNSShandbookDifferentialPositioning}.

To assess their predictability and other characteristics, we analyzed 281
adjustments associated with 12 unique Starlink v1.0 and v1.5 SVs and made the
following observations.

\subsubsection{Cadence}
Frame clock adjustment opportunities occur at an almost perfectly regular 1-Hz
cadence.  Of the 281 adjustments studied, all but one arrived within a few ms of
an integer second from the previous one, according to the frame clock $\tf$. The
single outlier, from a STARLINK-5666 (v1.5) capture, arrived 100 ms earlier than
expected.  At each opportunity, an adjustment may occur or not---note the lack
of adjustment at 6.5 s in Fig. \ref{fig:frameTimeSmoothingP}.

\subsubsection{Alignment with Respect to FAI}
One might expect frame clock adjustment opportunities to be aligned to the FAI
such that 15 inter-adjustment intervals fit neatly within one FAI.  This is not
the case; there appears to be no fixed relationship between FAI boundaries and
the 1-Hz adjustment opportunities.  On this observation and a supporting one
presented in Section \ref{sec:comp-sign-analys} rests our decision to model the
GNSS disciplining of the base oscillator as occurring prior to and independent
of any further adjustments made by the frame clock, as illustrated in
Fig. \ref{fig:clock-block-diag}.  Thus, the frame clock $\tf$ exhibits the 1-Hz
disciplining adjustments present in $\td$, but the adjustment amplitude and
timing are not coordinated with the frame timing.

\subsubsection{Probability Distribution of Adjustment Amplitudes}
\begin{figure}[t]
\centering
\includegraphics[width=9.2cm]
{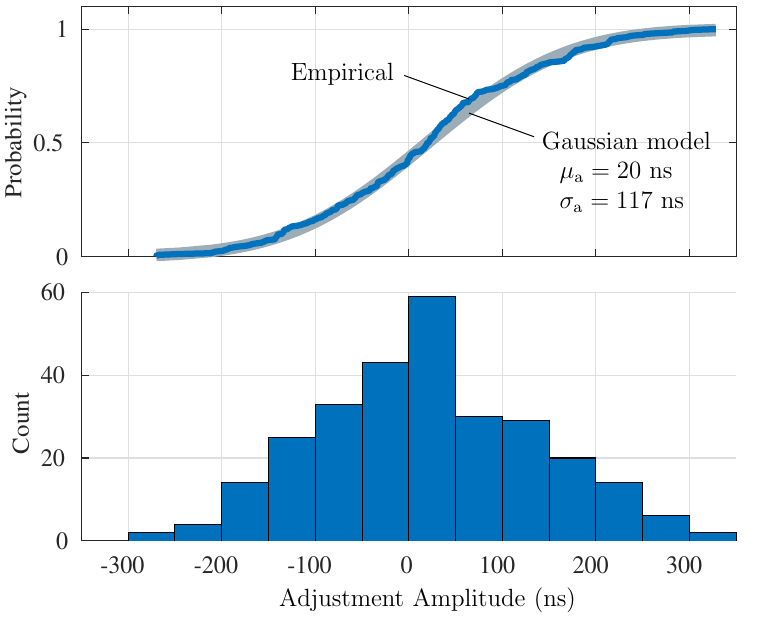}
\caption{Top: Cumulative probability distribution for Starlink v1.0 and v1.5
  frame clock adjustment amplitudes. Bottom: Histogram of adjustment amplitudes.}
\label{fig:jumpCdfP}
\end{figure}
If the 1-Hz frame clock adjustments were quantized (say, occurring only in 20-ns
steps), this would offer hope of developing an adjustment compensation mechanism
within a pseudorange-based position and time estimator.  Alas, this is not the
case.  Instead, the adjustment amplitudes appear to be smoothly distributed with
mean $\mu_\text{a} = 20$ ns and standard deviation $\sigma_\text{a} = 117$ ns,
as shown in Fig. \ref{fig:jumpCdfP}. A Shapiro-Wilk goodness-of-fit test for
normality yielded $p = 0.3$, suggesting a reasonably close alignment with the
Gaussian distribution \cite{royston1992approximating}.

With $\sigma_\text{a} = 117$ ns, the adjustment amplitudes are commensurate with
the symbol guard interval (the cyclic prefix) for Starlink's OFDM symbols, which
is $\Tg = 133$ ns \cite{humphreys2023starlinkSignalStructure}.  Thus, one might
expect the adjustments to cause inter-symbol interference (ISI) and thus a
degradation in communications throughput.  But close inspection reveals that the
adjustments are not instantaneous: they occur slowly relative to the OFDM symbol
rate, with a rise time of about 25 ms and a settling time of 150 ms.  Thus they
pose no risk of increased ISI.

\subsubsection{Frequency Adjustments}
For further insight into adjustment characteristics, we smooth $\ttrp$ by
eliminating the time shifts due to the 1-Hz adjustments.  To perform this
smoothing, we fit a 2nd-order polynomial to the inter-adjustment segments in
$\ttrp$, truncated to exclude the leading $0.15$ s and final $0.02$ s of data to
avoid the adjustments' transient effects.  The truncated inter-adjustment
polynomial fits are shown in green in the bottom panel of
Fig. \ref{fig:frameTimeSmoothingP}.  They and their corresponding data segments
are then shifted in such a way as to minimize first-order discontinuity.  That
is, for each inter-adjustment segment in the FAI, we minimize the difference
between the final value of the segment's shifted polynomial fit and the initial
value of the next segment's shifted fit.

The resulting time series, denoted $\ttrpp$ and exemplified by the black trace
in the bottom panel of Fig. \ref{fig:frameTimeSmoothingP}, reveals that the 1-Hz
disciplining adjustments not only impose a time shift in the base oscillator's
time $\td$ but also a frequency shift.  Note in
Fig. \ref{fig:frameTimeSmoothingP} the abrupt change in the slope of $\ttrpp$ at
the adjustment opportunity just prior to $\tf = 6$ s.  As with time shifts,
frequency shifts were found to occur randomly at each adjustment opportunity and
exhibit apparently random, unquantized amplitudes.  The presence of frequency
adjustments further complicates any effort at modeling $\dtf$ to obtain accurate
position and time estimates from pseudorange measurements.

\subsection{Nominal Jitter}
\label{sec:nominal-jitter}
\begin{figure}[t]
\centering
\includegraphics[width=\columnwidth]
{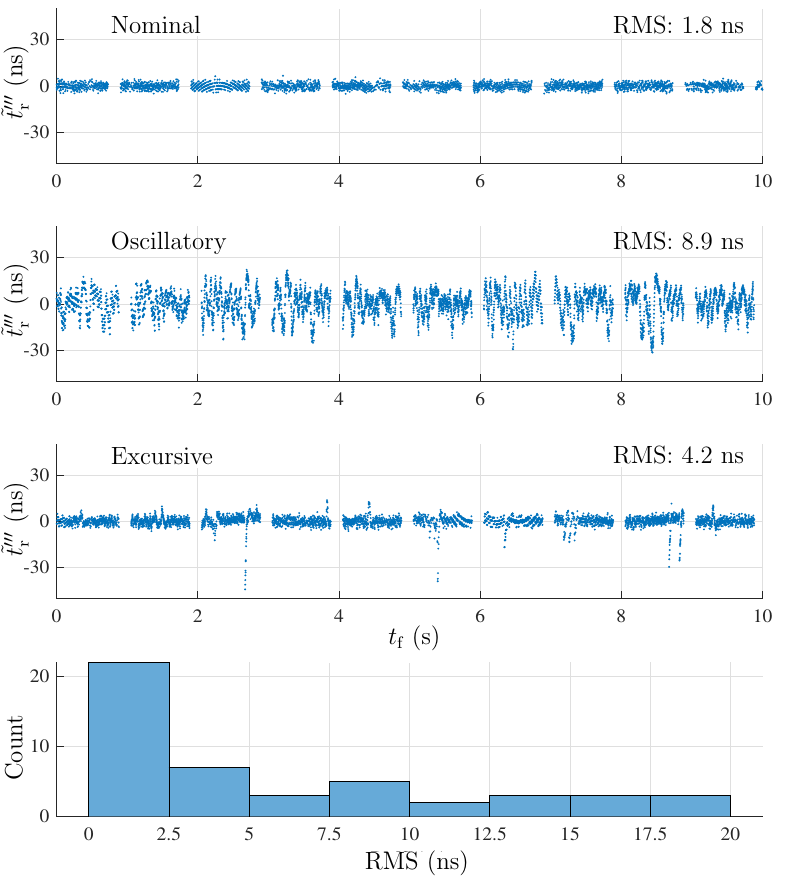}
\caption{Top three panels: The flattened time series $\ttrppp$ for the same
  Starlink SV during three contiguous FAIs, showing nominal, oscillatory, and
  excursive high-frequency frame clock behavior.  The data shown are for
  STARLINK-2119, a v1.0 SV, from signals captured in November 2023.  Bottom
  panel: Histogram of per-FAI $\ttrppp$ RMS values for all v1.0 and v1.5 SVs
  studied.  A single RMS measurement was extracted from the $\ttrppp$ time
  history corresponding to the dominant signal of each FAI.}
\label{fig:jitterSummaryP}
\end{figure}

A further processing step allows us to assess the high frequency variations
(jitter) in $\ttr$.  We subtract from $\ttrp$ the 2nd-order polynomial fits (the
green segments in Fig. \ref{fig:frameTimeSmoothingP}) of each truncated
inter-adjustment segment to flatten the time series.  We denote this flattened
time series as $\ttrppp$, an example of which is shown in the top plot of
Fig. \ref{fig:jitterSummaryP}.  Under nominal conditions for v1.0 and v1.5 SVs,
the RMS value of $\ttrppp$ ranges between 1.7 and 2.5 ns.  A large contributor
to this jitter is nearest-sample quantization noise, which, for $y_{m01}[k]$
from (\ref{eq:localReplicaWithDoppler}) sampled at $\Fs = 240$ MHz, has an RMS
value of $1/\sqrt{12}\Fs = 1.2$ ns \cite{gray1998quantization}.  Assuming this
noise is independent of other sources of jitter, the RMS contribution of the
other sources ranges from 1.5 to 2.2 ns.  Some of this is due to thermal noise
in the receiver, so 2.2 ns serves an an upper bound on the jitter in the SVs'
frame clock deviation $\dtf$ under nominal behavior.  This implies that the
frame clock of Starlink v1.0 and v1.5 SVs is capable of maintaining jitter at
the ns level, which, setting aside the 1-Hz time and frequency adjustments and
the lower-frequency variations in $\dtf$, is adequately low to support accurate
pseudorange-based PNT.

\subsection{Short-Term Frame Clock Stability Bound}
\label{sec:short-term-stability}
\begin{figure}[t]
\centering
\includegraphics[width=\columnwidth]
{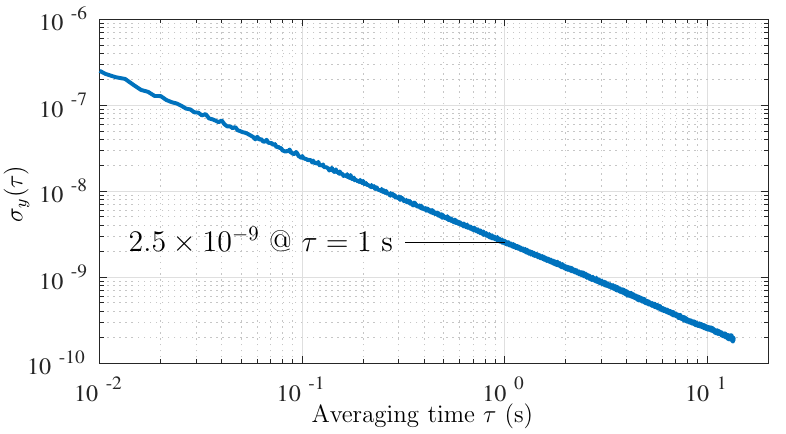}
\caption{Overlapping Allan deviation of a composite frame clock constructed from
  18 $\ttrpp$ time histories originating from 9 distinct v1.0 and v1.5 Starlink
  SVs.}
\label{fig:adevPlotP}
\end{figure}

To probe the stability limits of the v1.0 and v1.5 Starlink SV frame clocks, we
performed an Allan deviation analysis of $\ttrpp$ time histories, such as the
black trace in Fig. \ref{fig:frameTimeSmoothingP}, for 18 separate FAIs.  The
duration of these $\ttrpp$ time histories ranged from 10 to 15 s.  The data
originate from 9 unique v1.0 and v1.5 Starlink SVs whose signals were captured
during 2022 and 2023 and whose frame timing was derived from dominant signals
with nominal RMS levels (2.5 ns or below) in the corresponding $\ttrppp$ traces.
The resulting overlapping Allan deviation, shown in Fig. \ref{fig:adevPlotP},
reveals the average short-term stability of v1.0 and v1.5 frame clocks.  Because
frame slot occupancy is based on user demand, and therefore stochastic, the
Allan deviation is derived from irregularly spaced data, as in
\cite{hackman1996noise}.  But for every averaging time $\tau$ shown in
Fig. \ref{fig:adevPlotP}, there were at least 1500 samples from which to
estimate the corresponding Allan variance $\sigma^2_y(\tau)$, enough to yield
highly accurate variance estimates.

Given the processing and data selection involved in creating the $\ttrpp$ time
histories on which the composite Allan deviation plot shown in
Fig. \ref{fig:adevPlotP} is based, which tends only to remove variation, the
plot should be taken as a lower bound on the frame clock stability of v1.0 and
v1.5 Starlink SVs.  This best-case stability is broadly consistent with a
temperature-compensated crystal oscillator (TCXO). For example, at an averaging
time of $\tau = 1$ s, the fractional frequency deviation is
$\sigma_y(\tau) = 2.5 \times 10^{-9}$, which is what one would expect from an
average-quality TCXO.  Thus, we may conclude that the short-term stability of
the Starlink v1.0 and v1.5 frame clocks is no better than that of a TCXO,
though, as will be shown, it can episodically be much worse.  One should bear in
mind, however, that a TCXO-quality base oscillator would not preclude
highly accurate pseudorange-based PNT. It is true that all traditional GNSS SVs
employ highly stable atomic oscillators \cite{beardAndSeniorGNSShandbook}, but a
PNT service fused with a broadband LEO communications service could get by with
much cheaper and less stable clocks by sending near-zero-age clock models to
users through its high-capacity communications channels.  For example, assuming
the Allan deviation shown in Fig. \ref{fig:adevPlotP}, clock model updates at 1
Hz (corresponding to an averaging time of $\tau = 1$ s) would be sufficient to
keep satellite clock modeling errors in the PNT solution below about 2.5 ns.

\subsection{Anomalous Frame Clock Behavior}
\label{sub:anom-behavior}
The foregoing jitter and stability results apply only for nominal behavior of
the Starlink frame clock, which holds for approximately half of the FAIs
studied.  The other half manifest various modes of high-frequency frame clock
instability.  We do not know the underlying cause of these anomalous frame clock
behavior modes, but we note here certain patterns of behavior and offer
conjectures about their meaning.

Consider the top three panels of Fig. \ref{fig:jitterSummaryP}, which show the
flattened time series $\ttrppp$ for 10-s intervals within three separate FAIs.
The top panel shows nominal behavior, with 1.8 ns RMS, and acts as a baseline
for comparison.  The second panel shows much higher-RMS variations with
significant time correlation.  In this example trace, the deviations are bounded
between $-13$ ns and 28 ns and include strong frequency components at 7.2 and
13.7 Hz.  We classify anomalous frame clock behavior as oscillatory, as in this
example, when one or two spectral components dominate.  The third panel shows a
$\ttrppp$ time history with sudden and irregularly spaced deviations, e.g., at
the 2.6 s and 5.4 s marks.  We classify such behavior as excursive.

The bottom panel of Fig. \ref{fig:jitterSummaryP} shows a histogram of $\ttrppp$
RMS values for all v1.0 and v1.5 SVs studied.  Each sample contributing to the
histogram is a scalar RMS measurement extracted from the $\ttrppp$ time history
corresponding to the dominant signal of a single FAI.  Of the 48 FAIs studied,
\begin{itemize}
\item 22 exhibited nominal frame clock behavior, with RMS values below 2.5
  ns;
\item 19 exhibited oscillatory behavior, with RMS values above 2.5 ns and
  having a principal frequency component typically residing between 12 and 14
  Hz;
\item 3 exhibited excursive behavior;
\item and 4 exhibited other behaviors, such as a mix of oscillatory and
  excursive modes, or elevated RMS values but without sudden excursions or
  dominant frequency components.
\end{itemize}

A clue to the origin of anomalous frame clock behavior may be found in the
following remarkable observation: Whatever mode the frame clock
manifests---whether nominal, oscillatory, excursive, or otherwise---invariably
persists during a full FAI, but can switch to a different mode in the next FAI,
even for the same SV.  In fact, the traces in the top three panels of
Fig. \ref{fig:jitterSummaryP} are for the same Starlink SV during three
contiguous FAIs.  From this clue we conclude that anomalous behavior is
connected to satellite hardware or software configuration changes that occur at
FAI boundaries.  For example, it may be that each of the three traces in
Fig. \ref{fig:jitterSummaryP} comes from an assigned beam cast by a different
one of the serving Starlink SV's downlink phase arrays, of which each SV has
three.  If the three phased arrays are driven by separate clocks, each with its
own characteristic high-frequency behavior, this would explain the FAI-aligned
frame clock mode switching.  Alternatively, it may be that the baseband frame
assembly process is governed by parameters set in software that remain fixed
over each FAI, and that some parameter combinations lead to nominal frame clock
behavior whereas others lead to anomalous modes.

One might suspect that apparent frame clock mode switching is actually unrelated
to any clock but instead due to complications in the TOA measurement process.
For example, oscillatory measurement errors might be caused by the interaction
of assigned beams and side beams whose frame slots are closely aligned in
composite signals.  Relatedly, excursive measurement errors could be caused by
the TOA measurement process occasionally misidentifying a side peak of the PSS +
SSS autocorrelation function as the primary peak in low SNR conditions.  Both of
these possibilities have been investigated and discarded.  In fact, anomalous
frame clock modes can manifest even in simplex signals (no significant side beam
peaks) with extremely high SNR.  For example, an FAI captured from STARLINK-3894
(v1.5) showed a mix of oscillatory and excursive frame clock behavior with
$\ttrppp$ RMS equal to 15 ns despite being an extraordinarily clean simplex
signal with post-correlation SNR of 41 dB. Conversely, composite signals (one or
more significant side beam signals), even with relatively low SNR, routinely
yield $\ttrppp$ RMS below 2.5 ns.  Clearly, the anomalous frame clock behavior
is inherent in the frame clock deviation $\dtf$ and not in any oddity of the TOA
measurement process.

As with the 1-Hz clock adjustments, the anomalous frame clock variations
discussed here would tend to degrade pseudorange-based PNT solutions formed
from frame TOA measurements.  While the root cause of anomalous frame clock
behavior remains a mystery, we emphasize that nearly half of the FAIs studied
showed nominal frame clock jitter, with RMS values below 2.5 ns.  Clearly, the
Starlink v1.0 and v1.5 baseband frame assembly process, and at least a large
fraction of its base oscillators, have sufficient stability to support
high-accuracy pseudorange-based PNT, provided the 1-Hz frame clock adjustments
could somehow be modeled or eliminated.

\subsection{Starlink v2.0 Frame Clock}
\begin{figure}[t]
\centering
\includegraphics[width=\columnwidth]
{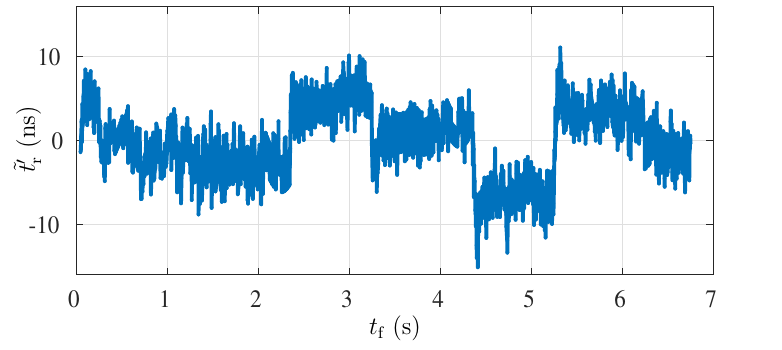}
\caption{As the top panel from Fig. \ref{fig:frameTimeSmoothingP} but for a
  v2.0-Mini SV. The data shown are for STARLINK-30178 from signals captured in
  February 2024.}
\label{fig:frameTime2p0P}
\end{figure}

We also generated $\ttrp$ time histories for v2.0-Mini SVs, an example of which
is shown in Fig \ref{fig:frameTime2p0P}.  These revealed frame clock adjustment
behavior differing from that of v1.0 and v1.5 SVs in two ways: (1) the
adjustments' magnitudes were generally much smaller, and (2) the adjustments
occurred at irregular intervals, as opposed to the regular 1-Hz intervals for
v1.0 and v1.5 SVs.  Apparently, the Starlink v2.0-Mini SVs employ a different
mechanism for base oscillator GNSS disciplining.

For v2.0-Mini SVs, the RMS values of the flattened time series $\ttrppp$ are
higher than for the v1.0 and v1.5 SVs, ranging from 2.4 to 13.3 ns.  There are
three possible explanations for this: as compared to the v1.0 and v1.5 SVs,
the v2.0-Mini SVs could have (1) oscillators with worse high-frequency
stability, (2) a higher-noise GNSS disciplining technique, or (3) a frame
clock that adds more high-frequency noise to the base oscillator signal $\td$.
Even still, the high-frequency jitter in the v2.0-Mini $\dtf$ would not
preclude meter-accurate pseudorange-based positioning because a batch of
consecutive frame TOAs could be averaged to mitigate the higher jitter.

\section{Absolute Frame Timing}
\label{sec:absol-frame-timing}
Evidence of base clock GNSS disciplining raises the possibility that Starlink
frames are continuously steered toward alignment with GPST.  If true, this would
be a notable discovery, inviting immediate use of Starlink for opportunistic
pseudorange-based PNT.  One could assume $\dtf(t) = 0$ for both assigned and
side beams, which would be accurate to within the 1-Hz adjustment amplitude
standard deviation, $\sigma_\text{a} = 117$ ns, or about 35 m in equivalent
distance.  Combining this satellite clock error model with publicly accessible
Starlink ephemerides that are accurate (at least periodically) to within about
$\sigma_\text{e} = 10$ m \cite{liu2024maneuver}, one could form pseudoranges
from a sequence of measured frame TOAs and process these with standard GNSS
algorithms to obtain a position and time solution. Single-epoch 95\% horizontal
solution accuracy in this context can be approximated in meters as
$\sqrt{4\cdot\text{HDOP} \left(c^2 \sigma_\text{a}^2 +
    \sigma_\text{e}^2\right)}$ \cite{iannucci2020fused}, where $\text{HDOP}$ is
the horizontal dilution of precision factor and $c$ is the speed of light in
vacuum.  Given multiple signals from unique Starlink SVs with favorable
geometry, HDOP could be as low as 0.55 \cite{iannucci2020fused}, yielding a 95\%
horizontal error less than 54 m, tight enough for many applications of practical
interest.

But if the frame clock is not transparent ($\tf \neq \td$ in Fig.
\ref{fig:frameClockExamplePlotP}), then it may introduce errors that drive
frame timing far from alignment with GPST despite the base clock being GNSS
disciplined.  This could render errors based on the simplistic clock model
$\dtf(t) = 0$ too large for any useful PNT solution.  A more sophisticated
clock error model would be required.

This section addresses the question of whether Starlink frames are aligned
with GPST.  We first describe the methods used to deduce the time of
transmission (TOT) in GPST of the $m$th frame of the $l$th FAI, written
$t^*(l,m)$, given the corresponding frame TOA in GPST, $t_*(l,m)$.  We then
solve for $\dtf(l,m)$ and draw conclusions about Starlink frame clock steering
and FAI alignment to GPST. For this initial study, it will suffice to
determine $t^*(l,m)$ to better than $\sigma_\text{a} = 117$ ns.

\subsection{Time of Transmission Calculation}
\label{sec:frame-transm-time}
TOT calculation begins with the frame TOA measurement $\ttr(l,m)$ modeled in
(\ref{eq:peakMeasurementModel}).  We subtract from this an estimate of the
offset $\delta\tr(l,m)$ obtained from the simultaneously captured GNSS signals,
accounting for all cable delays, as described in Section
\ref{sec:signal-capture}.  This process allows us to determine $t_*(l,m)$ to
within a few ns, which, in turn, is related to $t^*(l,m)$ by
\begin{equation}
  \label{eq:tgt-txrx}
  t^*(l,m) = t_*(l,m) - \dTOF(l,m)
\end{equation}
What remains is to calculate the frame's time of flight, $\dTOF(l,m)$, modeled as
\begin{equation}
  \dTOF(l,m) = \sfrac{1}{c} \cdot \|\vb{r}_\text{r}(t_*(l,m)) -
  \vb{r}_\text{t}(t^*(l,m))\| + \datm \label{eq:tof_explicit}
\end{equation}
where $\vb{r}_\text{r}(t_*(l,m))$ is the receiver's location at the TOA,
$\vb{r}_\text{t}(t^*(l,m))$ is the transmitter's location at the TOT, and
$\datm$ is the atmospheric (neutral and ionospheric) delay experienced by the
signal over its path from transmitter to receiver. Substituting
(\ref{eq:tgt-txrx}) into (\ref{eq:tof_explicit}) yields the implicit
relationship
\begin{multline}
  \label{eq:tof_implicit}
  \dTOF(l,m) = \sfrac{1}{c} \cdot \|\vb{r}_\text{r}(t_*(l,m)) -
    \\  \vb{r}_\text{t}(t_*(l,m)-\dTOF(l,m))\| + \datm 
\end{multline}
from which $\dTOF(l,m)$ can readily be calculated numerically, provided $\datm$ 
and smooth transmitter and receiver location functions $\vb{r}_\text{t}(t)$ and
$\vb{r}_\text{r}(t)$.

To support determination of $t^*(l,m)$ to within $\sigma_\text{a} = 117$ ns,
errors in $\vb{r}_\text{t}(t)$, $\vb{r}_\text{r}(t)$, and $\datm$ must be small
relative to this amount.  For $\datm$, a Saastamoinen
\cite{j_saastamoinen72_ros} neutral atmospheric model with average surface
parameters was applied with Niell wet and dry mapping
functions\cite{a_niell96_arw}; ionospheric delays, which are minimal at Ku-band,
were ignored.  Errors in $\datm$ for all elevation angles are expected to be
less than 20 ns.  The receiver location $\vb{r}_\text{r}(t)$ was approximated as
constant, i.e., $\vb{r}_\text{r}(t) = \vb{r}_\text{r}$. Although the true
location, the antenna feedhorn's phase center, moves by up to one meter as the
dish assembly rotates to track SVs, omission of this movement from $\dTOF(l,m)$
calculations introduces negligible timing error compared to $\sigma_\text{a}$.

Determination of $\vb{r}_\text{t}(t)$ required more careful treatment.  While
public providers of orbital data such as Celestrak and SpaceTrack offer
regularly updated ephemerides for all operational Starlink SVs, neither source
reliably publishes these data continuously at the required precision. Most
open-access ephemeris providers, including Celestrak, publish data as two-line
elements, which typically exhibit km-level uncertainty at epoch. This is
insufficient: km-level position uncertainty corresponds to
\SI{}{\micro\second}-level timing uncertainty. SpaceTrack ephemerides offer
meter-level positioning for all Starlink SVs; however, such precision is only
available for a short time after epoch, with 1-$\sigma$ position uncertainty
climbing to 10 meters in less than one hour. This would have sufficed if the
data refresh epochs were predictable so that signal captures could be set to
coincide with ephemeris epochs. But SpaceTrack refreshes Starlink ephemerides up
to thrice a day with epochs that vary randomly by several hours. As such, use of
SpaceTrack datasets for $\vb{r}_\text{t}(t)$ was deemed impractical for the
purposes of this paper.

Instead, we developed our own ephemeris models based on raw observables produced
by the onboard GNSS receivers of a small number of v2.0-Mini SVs. These
observables, provided by SpaceX and processed through our GRID receiver's
central estimator as configured for LEO dynamics and a TCXO clock model, yielded
discrete position solutions with formal errors below 10 m (1-$\sigma$). Position
solutions were then fit over 12-13-min intervals with a 9th-order Chebyshev
polynomial to yield a smooth $\vb{r}_\text{t}(t)$. Fit residuals were below 0.5
m RMS per dimension.

\begin{figure}[t]
\centering
\includegraphics[width=\columnwidth]{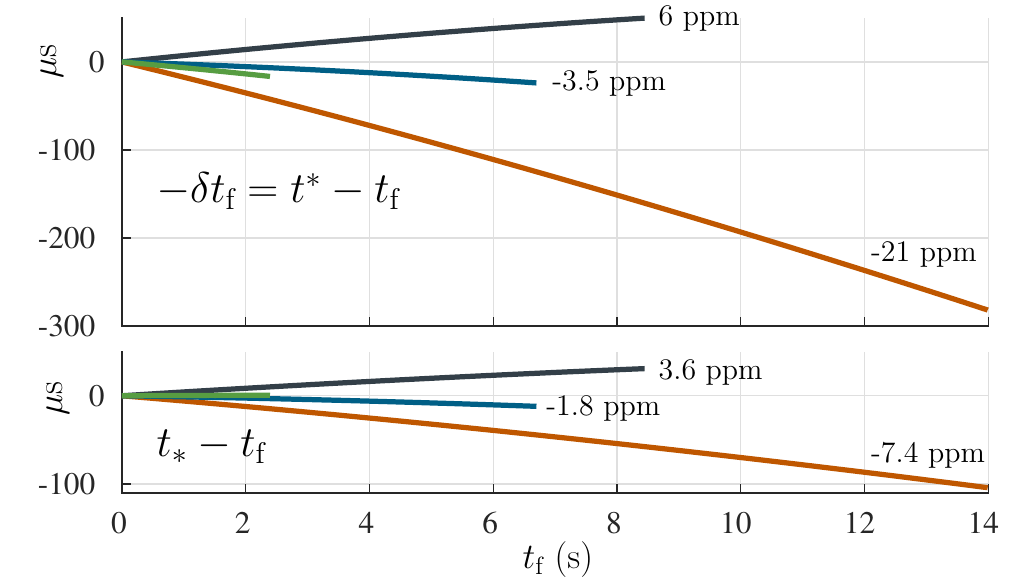}
\caption{Deviation of true frame TOT (top) and true frame TOA (bottom) from
  $\tf$ for subintervals of four representative FAIs from three different
  Starlink v2.0-Mini SVs.  For visual clarity, each trace has been shifted
  vertically to begin at an offset of zero. Colors indicate corresponding FAIs.
  Numerical labels indicate the clock drift in parts per million.}
\label{fig:frameClockDeviation}
\end{figure}

\subsection{Clock Steering}
We define clock steering as the process by which a clock is adjusted to match a
reference time such that the clock offset does not drift significantly over
relatively long ($>10$ s) durations.

The top panel in Fig. \ref{fig:frameClockDeviation} presents four frame clock
deviation traces $-\dtf(l,m) = t^*(l,m) - \tf(l,m)$. These traces exhibit clock
drift magnitudes varying from 0.2 to 21.4 ppm, values characteristic of a TCXO
at best and a basic quartz oscillator at worst. The bottom panel shows frame
clock deviations calculated from the same FAIs but substituting TOA in the place
of TOT.  We observe that all $t_*-\tf$ traces exhibit less deviation than their
corresponding $t^*-\tf$ traces in the top plot. This suggests that Starlink SVs
may be implementing frame clock pre-compensation to account for $\dTOF(l,m)$
changes as an SV passes overhead, thus mitigating the amount of frame clock
drift experienced by Starlink receivers. Nonetheless, with long-term drift
magnitudes ranging from 0.2 to 7.4 ppm, even the pre-compensated frame clock
drift is far from ideal for pseudorange-based PNT.

\begin{figure}[t]
\centering
\includegraphics[width=4.35cm]{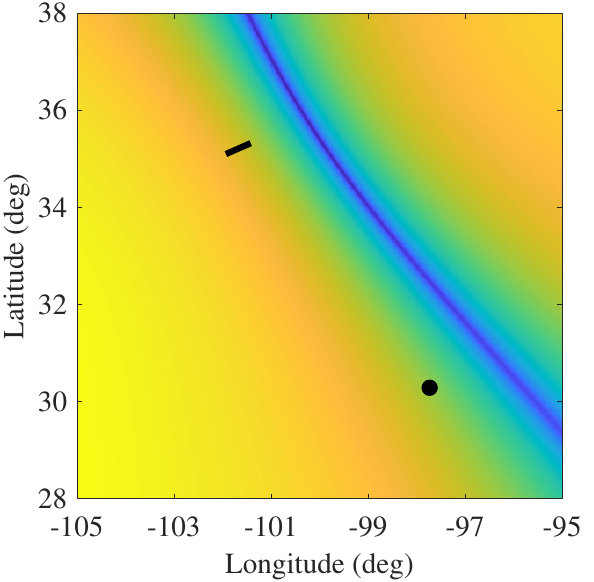}
\includegraphics[width=4.35cm]{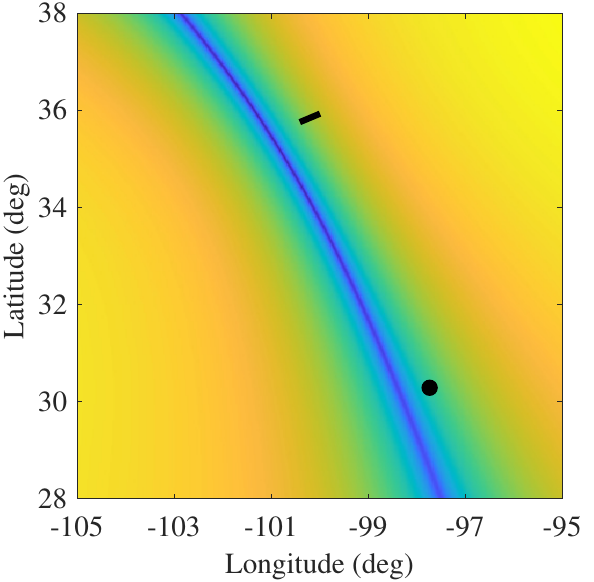}
\caption{Cost (on a dB scale) over the location search grid for ideal TOA clock
  steering. The left plot shows minimum-drift grid search results generated for
  the black line in the top panel of Fig. \ref{fig:frameClockDeviation}, whereas
  the right plot shows the same for the blue line. Both datasets originate from
  a capture taken from STARLINK-30178 in February 2024. Each plot's black circle
  represents the true receiver location. The black line shows the satellite
  ground track over the trace's time interval. Hot (yellow) areas signify areas
  experiencing degraded frame clock pre-compensation, while cold (blue) areas
  signify areas experiencing improved frame clock pre-compensation.}
\label{fig:flatlineCurves}
\end{figure}

It is worthwhile to consider the receiver location's impact on the effectiveness
of frame clock pre-compensation.  Pre-compensation cannot apply perfectly for
all receivers in a given service cell, since the evolution of $\dTOF(l,m)$
changes with $\vb{r}_\text{r}$. Possibly, a Starlink SV pre-compensates such
that receivers at the center of a service cell experience ideal clock steering
($t_*-\tf=0$).  To explore this possibility, we conducted a least-squares
location search on a grid spanning $10^\circ$ each of latitude and longitude
constrained to the surface of the WGS84 ellipsoid for two separate FAIs, those
corresponding to the black and blue traces of Fig.
\ref{fig:frameClockDeviation}. For each gridpoint, a trace was constructed as
\begin{equation}
  t_*(l,m) - \tf(l,m) = t^*(l,m) + \dTOFg(l,m) - \tf(l,m)
\end{equation} 
where $\dTOFg(l,m)$ is the time of flight from the transmitter to the
gridpoint's location (rather than our receiver's). The gridpoint was then
assigned a corresponding squared sum cost
$J = \sum_{m \in \Ml} [t_*(l,m) - \tf(l,m)]^2$, where $\Ml$ represents all
available occupied frame indices of the $l$th FAI. The results are shown in
Fig. \ref{fig:flatlineCurves}. Examination of the minimum-drift valley indicates
that even in the best cases, $t_*(l,m) - \tf(l,m)$ traces still exhibited 0.1
ppm of gross clock drift. While this is certainly an improvement, it remains
difficult to predict where the minimum-drift location will be. Consider that the
two minimum-drift search results are generated for the same SV but for different
FAIs, yet the minimum-drift valleys are far apart and differ in shape. Moreover,
they appear unconnected with the location of the service cell or the SV's ground
track.  We conclude that Starlink frame timing is not driven towards alignment
with GPST in a way that is both precise and predictable.

\subsection{Fixed Assignment Interval Timing}
\label{sec:fixed-assignm-interv}
\begin{figure}[t]
\centering
\includegraphics[width=9.2cm]
{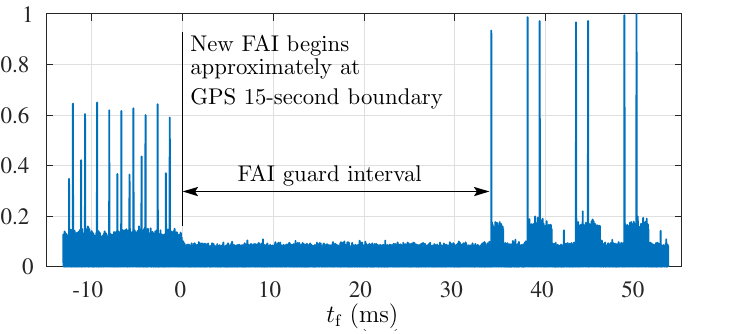}
\caption{Normalized cross-correlation of received Starlink data against a local
  PSS + SSS replica showing the boundary between subsequent FAIs.}
\label{fig:faiTransitionP}
\end{figure}

The foregoing subsection showed that, during each FAI, Starlink frame timing is
not driven toward alignment with GPST, whether at the SV or the target service
cell.  But it turns out that the frame clock does attempt approximate alignment
with GPST at the beginning of each FAI.  Fig. \ref{fig:faiTransitionP} shows the
boundary between two successive FAIs; call these the $(l-1)$th and $l$th FAIs.
Note that frame occupancy in the visible segment of the $(l-1)$th FAI appears
maximal.  Based on this, we assume that the final frame slot of the $(l-1)$th
FAI, whose index is $\Na - 1$, is occupied.

The $l$th FAI begins with the variable-length guard interval
$\Tag(l) = \Nag(l)\Tf$.  Analysis of a dozen such boundaries revealed that
$16 \leq \Nag(l) \leq 26$.  We then calculated the true TOT for the final frame
of the $(l-1)$th FAI, denoted $t^*(l-1,\Na(l) - 1)$, and estimated the true
start time of the $l$th FAI as $t^*(l,0) = t^*(l-1,\Na(l) - 1) + \Tf$.  For the
data shown in Fig. \ref{fig:faiTransitionP}, $t^*(l,0) = 514919.996874594$ when
expressed in GPS seconds of week.  Note that this is within about 3 ms of
514920, which is evenly divisible by 15.  A similar pattern was evident in all
such FAI transitions examined, from which we conclude that $t^*(l,0)$ ranges
between 4 and 2 ms before a 15-s GPST boundary.  This finding may be expressed
symbolically as 
\begin{align}
  \label{eq:FAIalignment}
  \lceil t^*(l,0) \rceil& -  t^*(l,0) \in [2, 4] ~ \mbox{ms} \\
  \lceil t^*(l,0) \rceil& \mod 15 = 0
\end{align}
where $\lceil \cdot \rceil$ denotes the ceiling function and $l \in \mathbb{N}$.

We also examined the true TOA $t_*(l,0)$ and found that, while closer to an
integer second, the range of variation was also approximately 2 ms, so it does
not appear that FAI boundary timing is being steered into precise alignment with
GPST at the location of the target service cell.

\section{Composite Signal Analysis}
\label{sec:comp-sign-analys}
All frame timing results presented so far have been for dominant signals,
typically from assigned beams.  But deeper insight into the Starlink timing
architecture can be gained by examining all constituent signals---both primary
and secondary---of composite signals. One might be curious to know, for example,
how closely secondary signals' frames are aligned with those of primary signals,
whether they experience the same 1-Hz adjustments and excursions, and whether
primary and secondary signals manifest similar levels of jitter.  The answers to
such questions would indicate whether the clock cascade in
Fig. \ref{fig:clock-block-diag} should be thought of as beam-specific,
phased-array-specific, or otherwise.

\begin{figure}[t]
\centering
\includegraphics[width=9.2cm]
{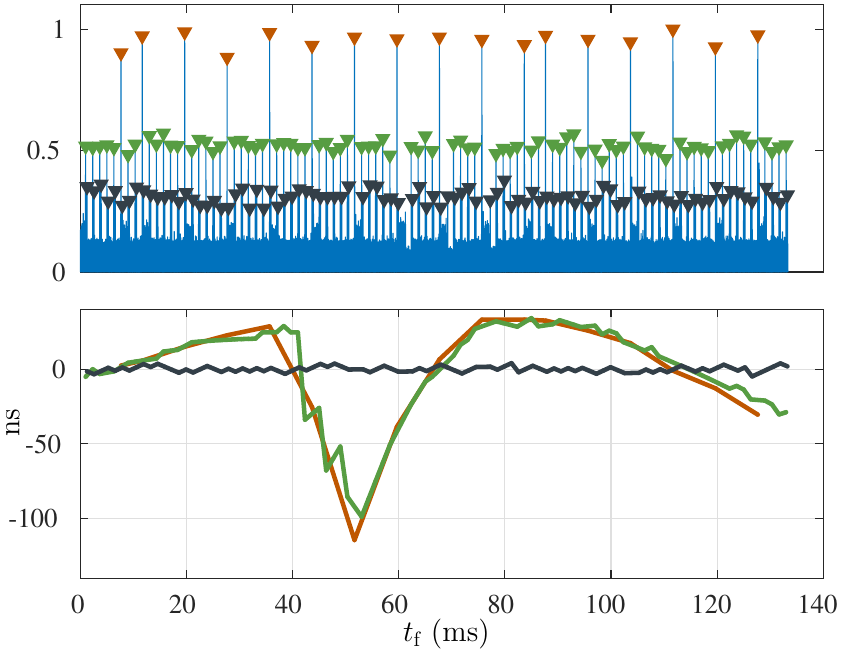}
\caption{Top: Normalized cross-correlation of received Starlink data against a
  local PSS + SSS replica over 100 frame intervals, showing peak trains from a
  dominant (probably assigned) beam (orange) and from two side beams (green and
  black).  Bottom: Corresponding measured frame TOA time histories $\ttr$ with a
  linear trend removed from each.  The data shown are for STARLINK-4577, a v1.5
  SV, from signals captured in November 2023.}
\label{fig:compositeSignalExampleP}
\end{figure}

The composite signal shown in the top panel of
Fig. \ref{fig:compositeSignalExampleP} is an informative example in this regard.
It consists of a dominant signal, marked in orange, and two clear secondary
signals, marked in green and black.  In the interval shown, the dominant signal
experiences a large (150-ns) excursion, evident in the orange trace of the lower
panel.  Notably, one of the secondary signals exhibits nearly the same excursion
whereas the other does not.  What is more, just 250 ms before the interval shown
there occurred a 1-Hz adjustment of the type discussed in Section
\ref{sub:nom-behavior}.  Different from the excursion, this adjustment affected
all three signals identically.  All composite signals we studied followed the
same pattern: excursions and oscillations may differ among beams of the same SV,
but adjustments do not.  This pattern implies a clock cascade model in which a
single base oscillator underlies all transmissions from a given SV, but separate
frame clocks drive disjoint subsets of the SV's beams.

For each signal in Fig. \ref{fig:compositeSignalExampleP}, Table
\ref{tab:SignalStats} gives the number of frames whose TOAs were measured, the
post-correlation SNR, the frame phase offset relative to the dominant signal's
phase, the RMS of the detrended traces in the bottom panel of
Fig. \ref{fig:compositeSignalExampleP}, and the frame rate offset parameter
$\betaf$.  The latter is a measure of how compressed or stretched the signal's
frame TOAs are compared to an exact $\Ff = 750$ Hz cadence.  The values for
$\betaf$ are derived from a least-squares fit to the TOA time history for each
signal.  For the signals marked in orange and green, the excursion was removed
before performing the least-squares fit.  The uncertainty range shown for
$\betaf$ is $\pm \sigma_\text{f}$, where $\sigma_\text{f}^2$ is the
least-squares error variance for $\betaf$ assuming TOA measurement noise
deviation $\sigma_w = 2$ ns.

\begin{table}[t]
  \centering
  \caption{Signal Statistics for Fig. \ref{fig:compositeSignalExampleP}}
  \begin{tabular}[c]{lcccrl}
    \toprule
    Signal &  Frames & SNR & Offset  & RMS  & ~~~~$\betaf \times 10^5$ \\
           &         & (dB)&  (ns)   & (ns)~ & \\ \midrule                                  
    Orange & 17 & 28 & 0 & 37.2 & $-1.2 \pm 0.0014$ \\ 
    Green  & 80 & 23 & 4.3 & 31.2 & $-1.2 \pm 0.0006$ \\ 
    Black  & 95 & 20 & $2.6 \times 10^5$ & 1.8 & $-1.1 \pm 0.0005$ \\ \bottomrule
  \end{tabular}
  \label{tab:SignalStats}  
\end{table}

The quantity $\betaf$ is the analog of the CFO parameter $\beta$ but for the
frame rate rather than the carrier frequency.  For all traditional GNSS,
$\betaf = \beta$ for received signals (to within minor ionosphere-induced
code-carrier divergence) because the same clock drives both carrier and
spreading code generation.  The local PSS + SSS replica in
(\ref{eq:coherentCombPssSss}) also assumes identical $\beta$ for both the
carrier and the modulation $x_{m01}(t)$.  But Table \ref{tab:SignalStats}
identifies $\betaf$ as distinct from $\beta$ because we discovered that
$\betaf = \beta$ does not hold universally for Starlink signals.

Table \ref{tab:SignalStats} indicates that the orange and green signals, which
both experience the excursion, are also closely tied in other ways: their frame
intervals are nearly perfectly aligned (within 4.3 ms), and they have identical
$\betaf$ values.  Contrarily, the excursion-free signal has a frame phase far
from the other two (amounting to an offset of nearly 20\% of the frame interval
$\Tf$) and a significantly different $\betaf$ value.  For all three signals, the
CFO parameter $\beta = (1.06 \pm 0.025) \times 10^{-5}$, where the uncertainty
range comes from Fig. \ref{fig:doppler_bound} for -10 dB pre-correlation SNR, a
conservative bound for the signals in Fig. \ref{fig:compositeSignalExampleP}.
Thus, in this case as in many others studied, $\betaf$ and $\beta$ are
significantly different---even different in sign.  More generally, we found that
$\betaf = \beta$ only holds for about half of the FAIs studied, and that the
condition $\betaf = \beta$ can change at FAI boundaries for the same SV.  This
supports a clock cascade model in which $\Df(t)$ and $\Dc(t)$ in
(\ref{eq:frameOffsetFrom_td}) and (\ref{eq:carrierOffsetFrom_td}) are
independent.



%

\section{Conclusions}
Based on our study of Starlink v1.0, v1.5, and v2.0-Mini frame timing, we draw
the following conclusions (C) and make the following conjectures (G) about
Starlink's timing architecture.
\newcounter{counter} 
\begin{list}{C\arabic{counter}}
  {\usecounter{counter}\leftmargin=1.5em}
\item Beam-to-cell assignments remain static over 15-s fixed assignment
  intervals (FAIs). Each downlink beam carries a sequence of frames containing
  user data.
\item All beams cast by a given SV carry signals that are ultimately driven
  by the same base oscillator, but the signals may or may not share the same
  frame clock, which means that frame phase, rate, and short-term stability can
  differ significantly from beam to beam.
\item Periodic adjustments of the base oscillator cause large ($\sim120$ ns
  for v1.0 and v1.5 SVs, $\sim10$ ns for v2.0-Mini SVs) abrupt deviations
  in frame timing that are common across all beams of the same SV.
\item Frame TOA jitter is nominally small (below 2.5 ns RMS), but frame timing
  can episodically manifest large unpredictable oscillations and excursions.
\item Frame TOA behavior can change on FAI boundaries, even for the same SV.
\item Frame short-term timing stability is ultimately limited by the quality
  of the base oscillator, which may be as good as, but no better than, a TCXO.
\item Frame timing is not driven toward alignment with GPST, whether at the
  SV or the target service cell.  Rather, frames as transmitted can have a rate
  that is significantly different (e.g., 20 ppm) from the nominal $\Ff = 750$ Hz
  rate.
\item FAIs are aligned within a few ms to a 15-s boundary of GPS seconds of
  week.
\item The carrier phase offset parameter $\beta$ and frame rate offset
  parameter $\betaf$ may differ significantly, unlike with traditional GNSS
  signals, for which $\betaf = \beta$ holds universally (to within minor
  ionosphere-induced code-carrier divergence).
\end{list}
\begin{list}{G\arabic{counter}}
  {\usecounter{counter}\leftmargin=1.5em}
\item The periodic adjustments seen in frame timing are the result of
  GNSS disciplining of the base oscillator.
\item The carrier signal is driven directly by the base oscillator, meaning that
  the carrier clock in Fig. \ref{fig:clock-block-diag} is transparent.  By
  contrast, frames are assembled at baseband and then mixed to RF and modulated
  onto the carrier by a process that is subject to the variations (including
  those due to GNSS disciplining) of the base oscillator but that may impose
  additional frame timing and rate variations.
\item Signals having nearly identical frame rate offset parameter $\betaf$
  and frame phase offset share the same frame clock.
\item Signals sharing the same frame clock are carried by beams emanating
  from the same downlink phased array, of which each Starlink SV has three.
\item Abnormal frame clock behaviors (oscillations, excursions) are not
  intentional.  They are the result of frame packaging and modulation algorithms
  that satisfy communications requirements but take no special care to ensure
  frame timing stability.
\item The frame packaging and modulation algorithms are software-defined and
  could be updated to eliminate excursions and oscillations.
\end{list}

Three key implications (I) for pseudorange-based positioning and timing
based on Starlink follow from the foregoing conclusions and conjectures:
\begin{list}{I\arabic{counter}}
  {\usecounter{counter}\leftmargin=1.5em}
\item Large and apparently unpredictable variations in frame timing, which can
  differ from beam to beam, and the lack of fine-grained frame phase steering
  toward a common time standard such as GPST, make Starlink currently unsuitable
  for purely opportunistic pseudorange-based PNT despite public availability of
  SV ephemerides.
\item Current Starlink SV hardware appears fundamentally capable of supporting
  pseudorange-based PNT with position and timing accuracy exceeding traditional
  GNSS.
\item There exist three options for making Starlink suitable for high-accuracy
  pseudorange-based PNT: (1) SpaceX could update Starlink's software-defined
  frame clock to eliminate abnormal oscillations and excursions and to steer
  frames into precise (ns-level) alignment with GPST at the transmitter; (2)
  SpaceX could broadcast a low-latency frame clock model $\dtf(t)$ that would
  allow users to compensate for frame clock variations; (3) a third party could
  establish a network of reference stations that would measure Starlink frame
  TOA and send a low-latency frame clock model $\dtf(t)$ to its subscribers.
\end{list}
We note that option (2) could be limited to Starlink subscribers, whereas (1)
and (3) would allow non-subscribers to benefit from Starlink-based PNT.  A
subscriber-limited version of (2) would involve bi-directional communication (at
least for subscriber authentication), whereas a subscription-free version would
enable passive (radio silent) Starlink-powered PNT.  Option (3) would require a
reference network dense enough to obtain accurate models $\dtf(t)$ for all
unique frame clocks.  In the worst case, this would require one reference
receiver per $\sim$20-km-diameter service cell.

\section*{Acknowledgments}
Research was supported in part by the by the U.S. Department of Transportation
under Grant 69A3552348327 for the CARMEN+ University Transportation Center, by
Sandia National Laboratories, and by affiliates of the 6G@UT center within the
Wireless Networking and Communications Group at The University of Texas at
Austin. Sandia National Laboratories is a multi-mission laboratory managed and
operated by National Technology \& Engineering Solutions of Sandia, LLC, a
wholly owned subsidiary of Honeywell International Inc., for the U.S.
Department of Energy’s National Nuclear Security Administration under contract
DE-NA0003525. This paper describes objective technical results and analysis.
Any subjective views or opinions that might be expressed in the paper do not
necessarily represent the views of the U.S. Department of Energy or the United
States Government.  The United States Government retains a non-exclusive,
paid-up, irrevocable, world-wide license to publish or reproduce the published
form of this article or allow others to do so, for United States Government
purposes. 

\bibliographystyle{ieeetran} 
\bibliography{pangea}
\end{document}